\documentclass[twocolumn]{aastex61}
\usepackage{amsmath}

\usepackage{color}
\usepackage{natbib}
\defcitealias{marble10}{M10}
\submitjournal{ApJ}

\shorttitle{HCO$^{+}$ and HCN emission in N\,55}
\shortauthors{Nayana et al.}

\begin{document}

\title{ALMA observations of HCO$^{+}$ and HCN emission in a massive star forming region N\,55 of the Large Magellanic Cloud}
\correspondingauthor{Naslim Neelamkodan} 
\email{naslim.n@uaeu.ac.ae}

\author[0000-0002-0844-6563]{Nayana~A.J.}
\affiliation{Department of physics, United Arab Emirates University, Al-Ain, UAE, 15551.}
%\affiliation{National Centre for Radio Astrophysics, Tata Institute of Fundamental Research, PO Box 3, Pune, 411007, India}
\author[0000-0001-8901-7287]{Naslim~N}
\affiliation{Department of physics, United Arab Emirates University, Al-Ain, UAE, 15551.}
\affiliation{Academia Sinica Institute of Astronomy and Astrophysics, Taipei 10617, Taiwan R.O.C}
\affiliation{Armagh Observatory, College Hill, Armagh, Northern Ireland, UK, BT61 9DG}
\author{T.~Onishi}
\affiliation{Department of Physical Science, Graduate School of Science, Osaka Prefecture University, 1-1 Gakuen-cho, Sakai, Osaka 599-8531, Japan}
\author[0000-0003-2743-8240]{F.~Kemper}
\affiliation{European Southern Observatory, Karl-Schwarzschild-Str. 2, 85748, Garching b. München, Germany}
\affiliation{Academia Sinica Institute of Astronomy and Astrophysics, Taipei 10617, Taiwan R.O.C}
\author[0000-0002-2062-1600]{K.~Tokuda}
\affiliation{Department of Physical Science, Graduate School of Science, Osaka Prefecture University, 1-1 Gakuen-cho, Sakai, Osaka 599-8531, Japan}
\author[0000-0003-3229-2899]{S.~C.~Madden}
\affiliation{Laboratoire AIM, CEA/DSM - CEA Saclay, 91191 Gif-sur-Yvette, France}
\author[0000-0002-5908-9543]{O.~Morata}
\affiliation{European Southern Observatory, Karl-Schwarzschild-Str. 2, 85748, Garching b. München, Germany}
\author{S.~Nasri}
\affiliation{Department of physics, United Arab Emirates University, Al-Ain, UAE, 15551.}
%\author[0000-0003-4358-0925]{S.~Hony}
%\affiliation{Universit\"{a}t Heidelberg, Zentrum f\"{u}r Astronomie, Institut
%f\"{u}r Theoretische Astrophysik, Albert-Ueberle-Str. 2, 69120 Heidelberg, Germany}

\author[0000-0002-0283-8689]{M.~Galametz}
\affiliation{Laboratoire AIM, CEA/DSM - CEA Saclay, 91191 Gif-sur-Yvette, France}
\begin{abstract}
We present the results of high spatial resolution HCO$^{+}$($1-0$) and HCN($1-0$) observations of N\,55 south region (N\,55-S) in the Large Magellanic Cloud (LMC), obtained with the Atacama Large Millimeter/submillimeter Array (ALMA). N\,55-S is a relatively less extreme star-forming region of the LMC characterized by a low radiation field. We carried out a detailed analysis of the molecular emission to investigate the relation between dense molecular clumps and star formation in the quiescent environment of N\,55-S. We detect ten molecular clumps with significant HCO$^{+}(1-0)$ emission and eight with significant HCN($1-0$) emission, and estimate the molecular clump masses by virial and local thermodynamic equilibrium analysis. All identified young stellar objects (YSOs) in the N\,55-S are found to be near the HCO$^{+}$ and HCN emission peaks showing the association of these clumps with recent star formation activity. The molecular clumps that have associated YSOs show relatively larger linewidths and masses than those without YSOs. We compare the clump properties of the N\,55-S with those of other giant molecular clouds (GMCs) in the LMC and find that N\,55-S clumps possess similar size but relatively lower linewidth and larger HCN/HCO$^{+}$(1$-$0) flux ratio. These results can be attributed to the low radiation field in N\,55-S resulted by relatively low star formation activity compared to other active star-forming regions like 30\,Doradus-10 and N\,159. The dense gas fraction of N\,55-S is $\sim$ 0.025, lower compared to other GMCs of the LMC supporting the low star formation efficiency of this region.

\end{abstract}

\section{Introduction}
\label{sec:introduction}
Massive stars form as clusters in the densest clumps of Giant Molecular Clouds (GMCs). These clumps are self-gravitating parsec sized structures that collapse and fragment into dense cores to form high-mass stars \citep{williams2000}. The physical processes that determine the fragmentation of GMCs to clumps and cores depend on the star formation history, cooling due to the build-up of metals and feedback from massive stars. Different star-forming environments can change the properties of GMCs and its sub-structure. Thus it is important to look at the GMCs and their sub-structure in different feedback environments to study star formation in galaxies.
 
Star formation studies on galaxy wide scale exist for the Milky Way \citep{kennicutt2012}. However, studying star formation in other galaxies is limited due to the lack of high-resolution instruments that can resolve the dense molecular clumps of GMCs. Such studies can be done in the Large Magellanic Cloud (LMC) due to its proximity. At a distance of 49.59\,kpc \citep{pietrzynski2019}, high spatial resolution observations of molecular clouds at sub-parsec scale can be obtained. Several massive star forming GMCs of the LMC has been studied extensively. These include 30-Doradus \citep{nayak2016,indebetouw2020}, N\,159 \citep{nayak2018,fukui2015,chen2010}, N\,44 \citep{chen2009,chen2010}, N\,113 \citep{wong2006,carlson2012} and N\,105 \citep{ambrocio1998,carlson2012}. The presence of stellar cluster R136 and numerous OB stars makes 30-Doradus the most extreme star forming GMC of the LMC \citep{nayak2016,hughes2010,pineda2009}. N\,159 is also a region of intense radiation field $\sim$ 156$\chi_{0}$ \citep[$\chi_{0}$ = 2.7$\times$10$^{-3} $ erg\,s$^{-1}$\,cm$^{-2}$;][]{draine1978}, high turbulence and shocks \citep{nayak2018,fukui2015}. With only 11 identified OB stars \citep{olsen2001} and a radiation field of 18$\chi_{0}$ \citep{pineda2009}, N\,55 is an example of quiescent star forming environment of the LMC.

\begin{figure*}
\begin{center}
\vspace{+0.9 cm}
\includegraphics[trim=0.2 0 0.4 0, clip, width=0.5\textwidth]{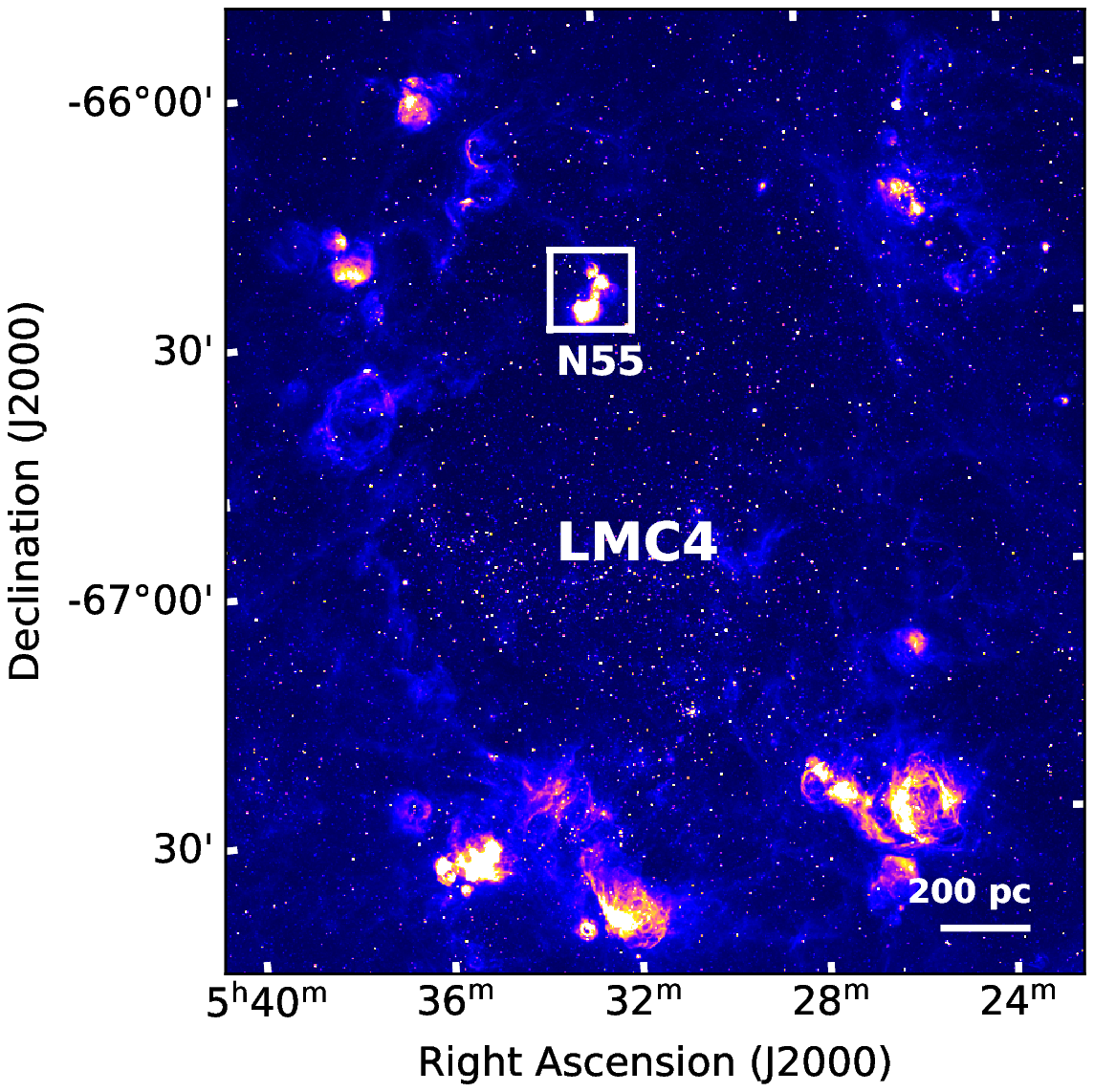}
\hspace{-0.2 cm}
\includegraphics[width=0.4\textwidth]{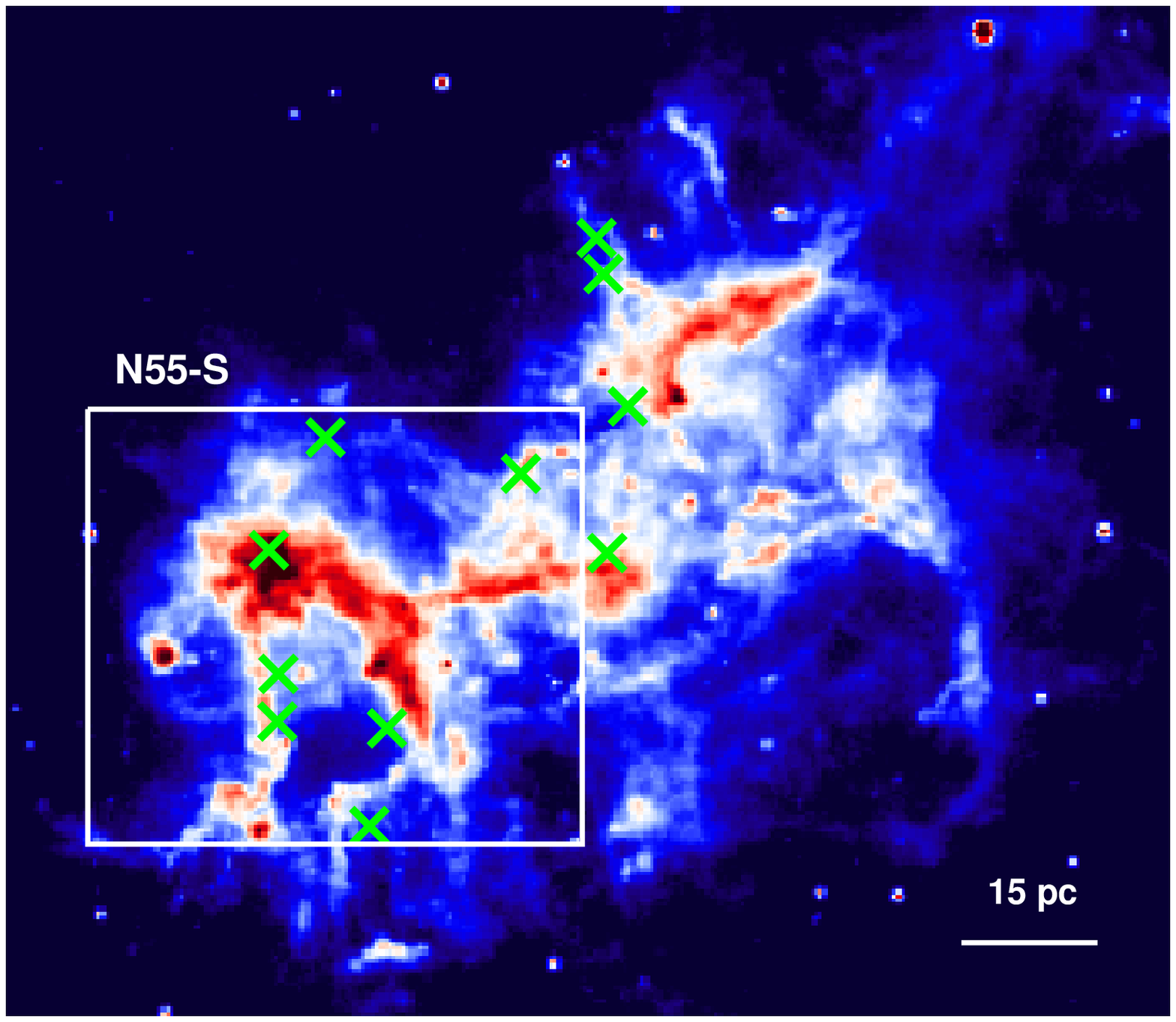}
\\
\vspace{+0.5 cm}
\caption {\scriptsize Left panel: The super-giant shell LMC\,4 in shown on an H${\alpha}$ image \citep{smith1998}. The location of N\,55 star forming region is marked. Right panel: The structure of N\,55 is shown in \textit{Spitzer} 8.0$\mu$m image \citep{meixner2010} which traces the filamentary PAH emission. The white box indicates the location of N\,55-S. The 11 OB stars are labelled in crosses \citep{olsen2001}.}
\label{fig:lmc}
\end{center}
\end{figure*}

N\,55 is a star forming region of size $\sim$ 60$\times$100 pc$^{2}$ located in the super-giant shell LMC\,4 \citep[see Fig \ref{fig:lmc};][]{yamaguchi2001}. While the bulk of LMC\,4 is empty of ionized gas, N\,55 stands out as a bright H\,{\sc ii} region in the H$\alpha$ map \citep{olsen2001}. The \textit{Spitzer} observations of N\,55 show filamentary distribution of Polycyclic Aromatic Hydrocarbon \citep[PAH; Fig \ref{fig:lmc};][]{naslim2018}. A total of 16 YSOs has been identified in the N\,55 from \textit{Spitzer} photometric observations indicating on-going star formation \citep{gruendl2009,seale2014}. \citet{naslim2015} detected the H$_2$ rotational transitions at 28.2 and 17.1\,$\micron$ in the N\,55 main molecular cloud complex with the infrared spectrograph onboard \textit{Spitzer} space telescope. Their studies shows a tight correlation of H$_2$ surface brightness with the PAH and total infrared emission indicating photoelectric heating caused by UV radiation from massive stars. The $^{12}$CO(1$-$0) and $^{13}$CO(1$-$0) observations of N\,55 reveals the clumpy nature of molecular gas with a total mass of $\sim$ 5.4 $\times$ 10$^{4}$ M$_{\odot}$ \citep{naslim2018}.  

CO isotopes probe molecular gas from relatively low-density regions ($\sim$10$^{2}$ cm$^{-3}$) whereas HCO$^{+}$(1$-$0) and HCN(1$-$0) trace dense molecular clumps due to their high critical densities ($\sim$10$^{4-5}$ cm$^{-3}$). In this paper, we present the HCO$^{+}$(1$-$0) and HCN(1$-$0) observations of the main molecular complex towards the south of the N\,55 (here onwards N\,55-S; see Fig \ref{fig:lmc}) with the Atacama Large Millimeter/submillimeter Array (ALMA). This is first time HCO$^{+}$ and HCN emission are observed toward N\,55. Our study aims to investigate the physical properties such as size, linewidth, mass, and dense gas fraction of the dense molecular clumps in a relatively quiescent region in the LMC. \cite{seale2012} studied molecular clumps of different star-forming GMCs of the LMC (N\,159, N\,105A, N\,44, and N\,113) using the same dense gas tracers. A similar study was carried out by \cite{anderson2014} in the 30\,Doradus-10. Our work extends this sample including N\,55-S with a goal to explore how different star formation environments affect the properties of dense molecular clumps. We compare the physical properties and scaling relation of the dense molecular clumps of the N\,55-S with other extreme star-forming environments, mainly 30-Doradus \citep{anderson2014} and N\,159 \citep{seale2012}.

The paper is organized as follows. We describe the ALMA observations and data analysis in \S \ref{sec:obs} and present the regions of molecular emission in \S \ref{sec:integrated-maps}. The clump identification and physical properties of each clump are presented in \S \ref{sec:clump-identification-characterization}. In \S \ref{sec:physical properties of dense molecular gas}, we inspect the spatial extent of dense gas tracers and CO isotopes to understand the density structure of the molecular cloud. We discuss the spatial coincidence of young stellar objects (YSOs) with the dense gas tracers and interpret the results in light of ongoing star formation in \S \ref{sec:Clump association with YSOs}. Finally, We compare the properties of the molecular clumps of the N\,55 cloud with other GMCs of the LMC in \S \ref{sec:comparison with other LMC clumps}. We summarize our results in \S \ref{sec:summary}. 

\section{ALMA Observations}
\label{sec:obs}
The ALMA observations of N\,55 main molecular complex were carried out in cycle 3 (project code 2013.1.00993.S) on 19 January 2015. The observations were done with the ALMA 12 m array in the band 3 in two frequency settings covering the HCN($1-0$) and HCO$^{+}$($1-0$) lines at rest frequencies 88.6318 and 89.1885 GHz, respectively. The field of view of the observation is $\sim$ 3.5 $\times$ 2 arcmin$^{2}$, centered at RA, Dec (J2000): 05:32:31.50; $-$66:26:22.5 covering $\sim$ 50 x 29 pc$^{2}$ in linear scale. Our observation is towards the southern region of N\,55 (N\,55-S) that covers the main molecular complex of N\,55 and does not cover the entire N\,55. The observations were carried out for a total integration time of 796 seconds. The correlator was set to have a bandwidth of 117.187 MHz split into 1920 channels in each spectral window. This corresponds to a native spectral resolution of 61 kHz (0.2 km\,s$^{-1}$). We binned the channels to a resolution of 0.4 km\,s$^{-1}$ for the analysis.

The data were reduced using common astronomy software application \citep[CASA;][]{mcmullin2007} package. Uranus was used as the flux calibrator and J\,0526$-$6749 was used as the phase calibrator. The ALMA pipeline calibrated visibilities were imaged using CASA task TCLEAN with a channel resolution of 0.4 km\,s$^{-1}$. We applied the Briggs weighting with a robust parameter of 0.5. The achieved rms sensitivities per 0.4 km\,s$^{-1}$ channel for HCO$^{+}$($1-0$) and HCN($1-0$) cubes are $\sim$ 10 and 11 mJy\,beam$^{-1}$ respectively.  The synthesized beam of HCO$^{+}$(1$-$0) map is 4.07$^{\prime\prime}$ $\times$ 3.11$^{\prime\prime}$ which translates to a linear size of 0.98 $\times$ 0.74 pc$^2$ at the LMC distance. For HCN(1$-$0) map, the synthesized beam is 4.13$^{\prime\prime}$ $\times$ 3.14$^{\prime\prime}$ which corresponds to 0.99 $\times$ 0.75 pc$^2$ in linear scale.  

\section{HCO and HCN emission in N\,55}
\label{sec:integrated-maps}
\begin{figure*}
\begin{center}
\vspace{+0.9 cm}
\includegraphics[trim=0.0cm 0cm 0.0cm 0cm, clip=true, scale=0.35]{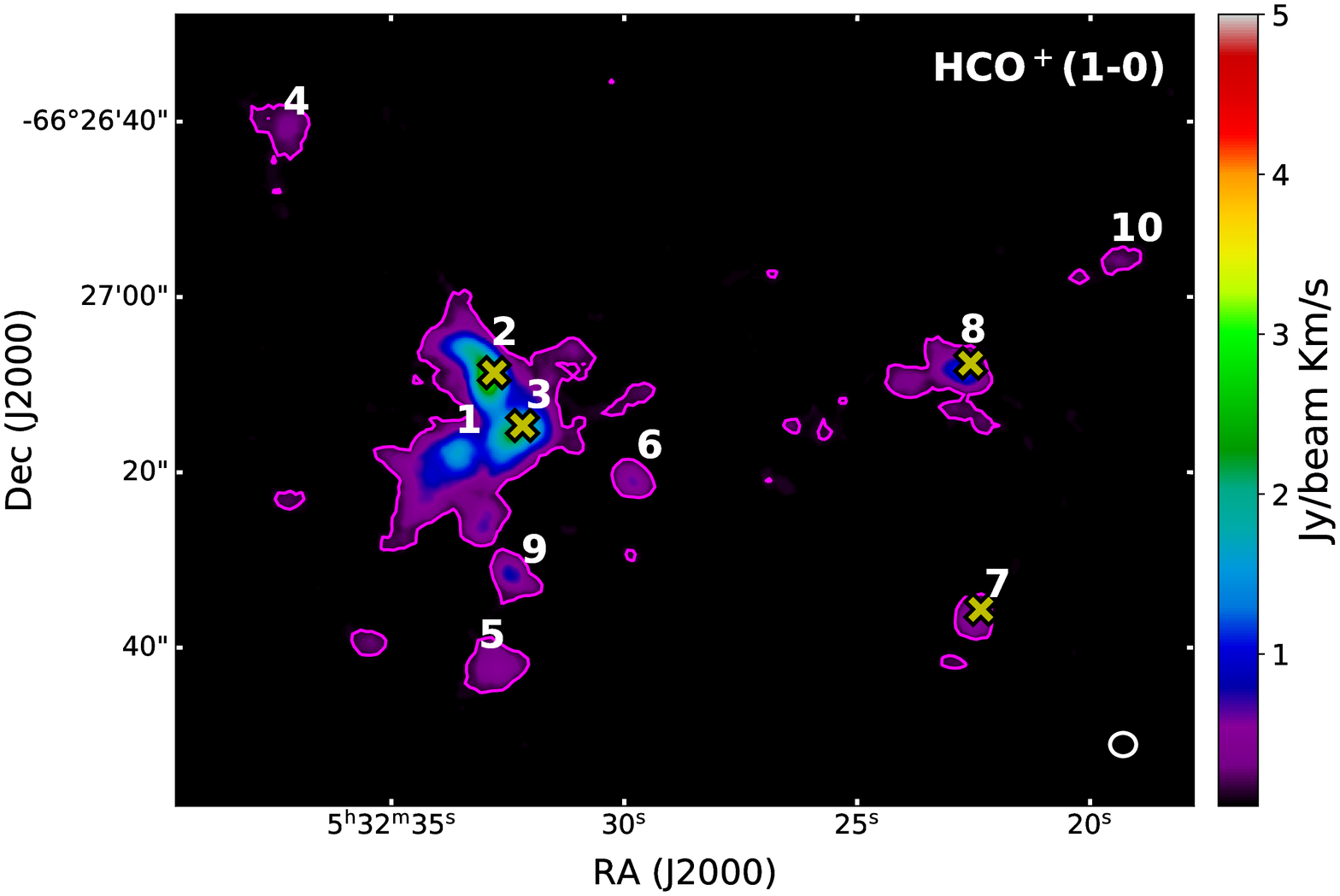}
%\vspace{-0.8 cm}
\hspace{-1.5 cm}
%\vspace{0.8 cm}
%\includegraphics[trim=0.0cm 0cm 0.0cm 0cm, clip=true, width=11cm, height=7.5cm]{HCN_clumps.pdf}
\includegraphics[trim=0.0cm 0cm 0.0cm 0.0cm, clip=true, scale=0.35]{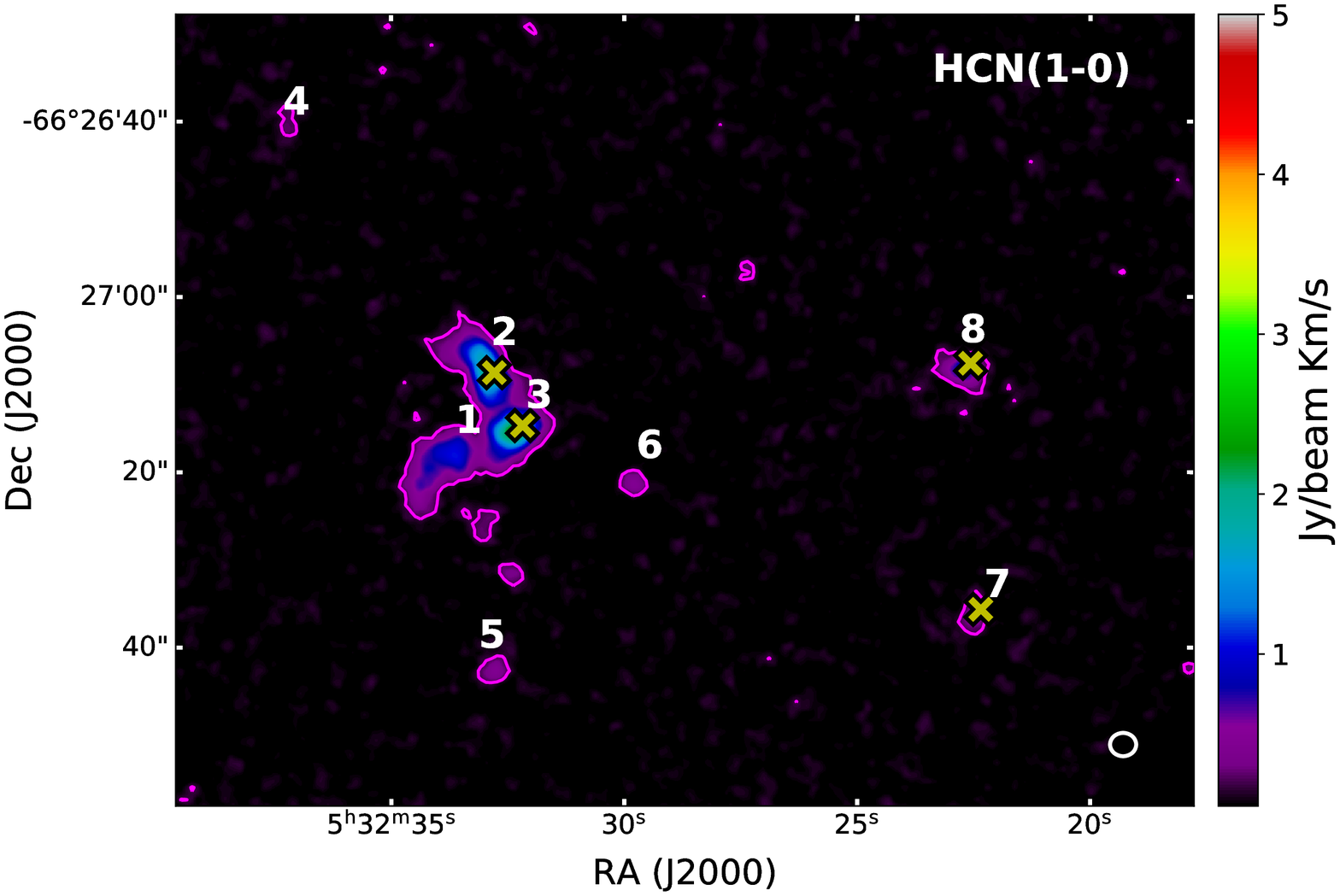}\\
\includegraphics[trim=0.0cm 0cm 0.0cm 0cm, clip=true, scale=0.35]{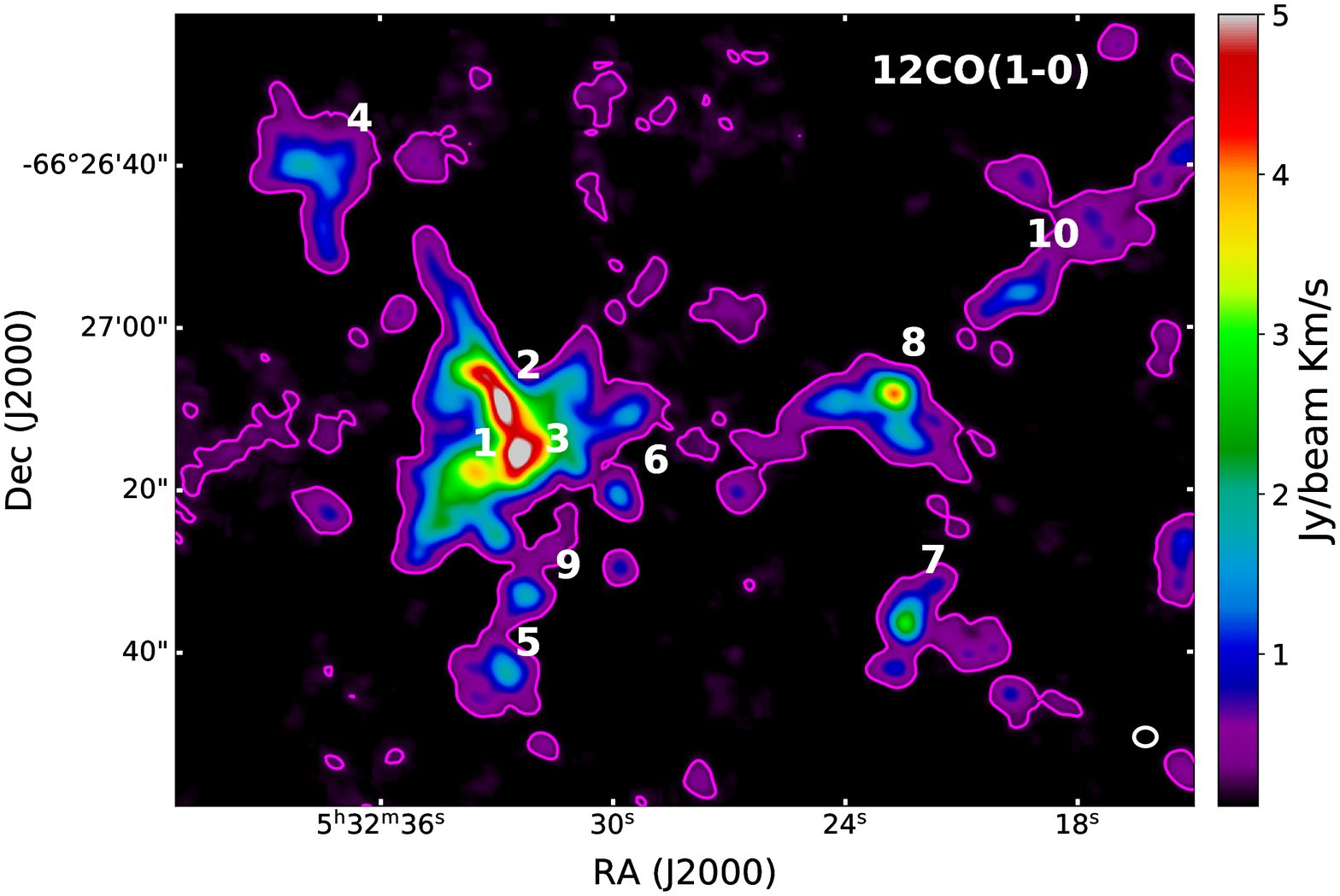}
\hspace{-1.5 cm}
\includegraphics[trim=0.0cm 0cm 0.0cm 0cm, clip=true, scale=0.35]{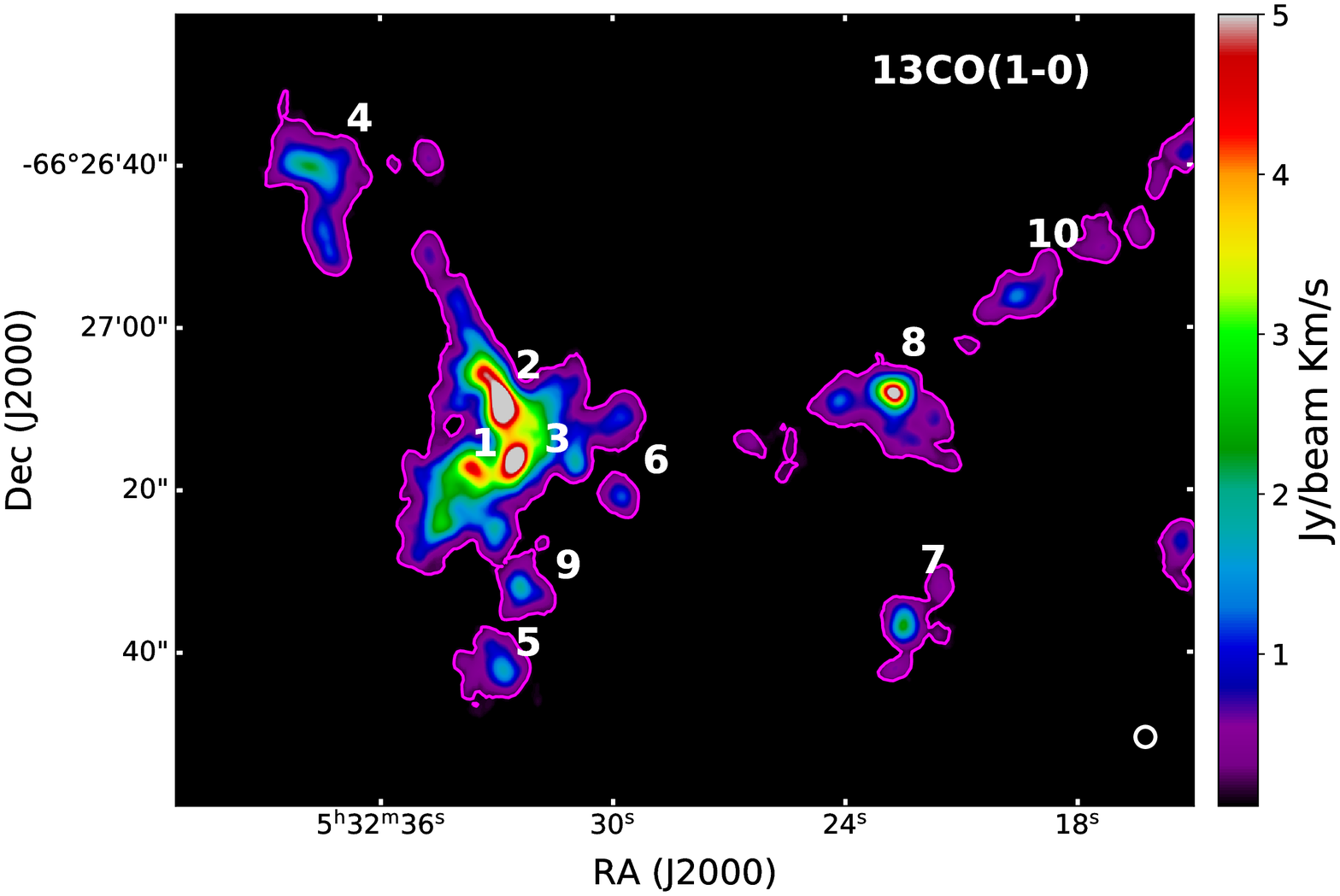}
\\
\vspace{+0.5 cm}
\caption {\scriptsize The velocity integrated intensity maps of HCO$^{+}$($1-0$) and HCN($1-0$) emission of the N\,55-S are shown along with $^{12}$CO($1-0$) and $^{13}$CO($1-0$) maps \citep{naslim2018} for comparison. 
The resolution of HCO$^{+}(1-0)$ and HCN$(1-0)$ maps are 4.07 $\times$ 3.11 arcsec$^{2}$ and 4.13 $\times$ 3.14 arcsec$^{2}$ respectively. 
The numbers denote the identified clumps and the yellow cross symbols denote the positions of YSOs  \citep{gruendl2009,chen2009}. The resolution of $^{12}$CO($1-0$) map is 3.53 $\times$ 2.32 arcsec$^{2}$ and that of $^{13}$CO($1-0$) map is 3.28 $\times$ 1.89 arcsec$^{2}$. The resolution elements are denoted in the bottom right corner of each map. }
\label{fig:all-clumps}
\end{center}
\end{figure*}

The velocity integrated intensity maps of HCO$^{+}$($1-0$) and HCN($1-0$) emission are shown in Fig \ref{fig:all-clumps}. These maps are obtained by integrating the emission line data cubes along the velocity axis. The maps show the clumpy nature of dense molecular gas in N\,55-S at parsec scales. The emission structures are mostly discrete, rather than nested or filamentary. A visual inspection suggests that the HCO$^{+}$($1-0$) emission from each molecular clump originates from slightly extended regions as compared to HCN($1-0$). This could be due to the higher densities probed by HCN($1-0$) emission as compared to HCO$^{+}$($1-0$). 
%\cite{olsen2001} identified total eleven B and O stars in the entire N\,55 which could be acting as the UV sources that {\red enhance the HCO$^{+}$ in the outer part of the clump}. 
A similar trend in the spatial distribution of HCO$^{+}$($1-0$) and HCN($1-0$) emission in molecular clumps has been reported in N\,105, N\,113, N\,159 and N\,44 regions \citep{seale2012}. The positions of YSOs in the N\,55-S region identified by \textit{Spitzer} observations  \citep{gruendl2009,chen2009} are shown along with HCO$^{+}$(1$-$0) and HCN(1$-$0) clumps in Fig \ref{fig:all-clumps}. The YSO positions are near the emission peaks of the clumps. The $^{12}$CO($1-0$) and $^{13}$CO($1-0$) maps of N\,55-S \citep{naslim2018} is also shown for a comparison which is discussed in \S \ref{sec:physical properties of dense molecular gas}.

\section{Clump identification and characterization}
\label{sec:clump-identification-characterization}
We use \textbf{astrodendro} \footnote{http://www.dendrograms.org} (a python package) to identify emission structures from the data cube. The algorithm identifies the hierarchical structure of emission \citep{rosolowsky2008}. Local maxima are identified from the data cube each with flux $>$3$\sigma_{\rm rms}$ and the iso-surfaces around the maxima are classified as leaves, branches or trunks.
If the iso-surfaces do not have any sub-structures, they are categorized as leaves. The largest contiguous structures in the cube are identified as trunks. The structures intermediate to leaves and trunks are classified as branches. Thus in a dendrogram, leaves do not overlap each other and they are the smallest emission clumps without sub-structures.

\textbf{Astrodendro} determines the basic parameters of the identified structures using the bisection method. The parameters are the velocity and positional centroids, integrated flux density, velocity dispersion (ie. rms linewidth $\sigma_{\rm v}$), rms sizes of the major ($\sigma_{\rm maj}$) and minor ($\sigma_{\rm min}$) axes of the clump and the position angle of the major axis. For Gaussian line profiles, the FWHM linewidth ($\Delta v$) is given by $\Delta v$ = $\sqrt{8ln(2)}$ $\sigma_{v}$. The spherical radius of the clump is $R=1.91\sigma_{r}$ where $\sigma_{\rm r}=\sqrt{\sigma_{\rm maj}\sigma_{\rm min}}$. We use bootstrapping technique to determine the uncertainties in the derived parameters.

We identify 10 significant molecular clumps in the HCO$^{+}$(1$-$0) and 8 in the HCN(1$-$0) data cubes. These are all identified as leaves by \textbf{astrodendro}. The properties of all these structures are listed in Table \ref{tab:clump-details-astrodendro}. For the resolution of our observations, we can resolve only clumps of size $\geq$4 arcsec. This translates to $\sim$ 1 pc at the LMC distance. The sizes of identified clumps are $1-2.2$ pc, except for one clump (clump id = 6) with size 0.82 pc. The sizes of dense molecular clumps in the massive star-forming regions of the Milky Way Galaxy is $\sim$ $0.1-1$ pc \citep{romero2017}. Given the limit in our spatial resolution, the smallest detected molecular structures in the N\,55-S seem compatible with the Milky Way clumps.

\begin{figure*}
\begin{center}
\hspace{0.0 cm}
\includegraphics[scale=0.30]{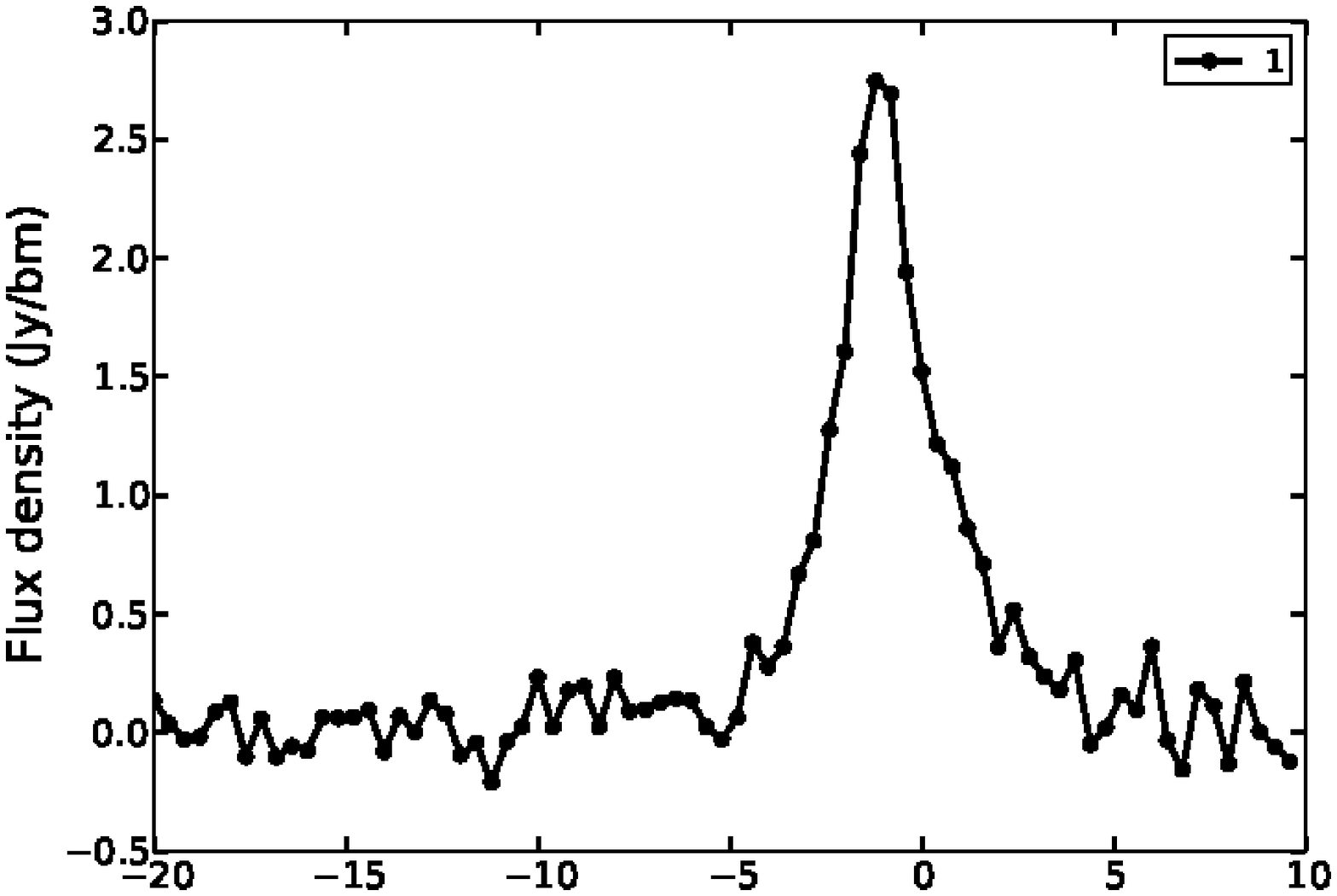}
\includegraphics[scale=0.30]{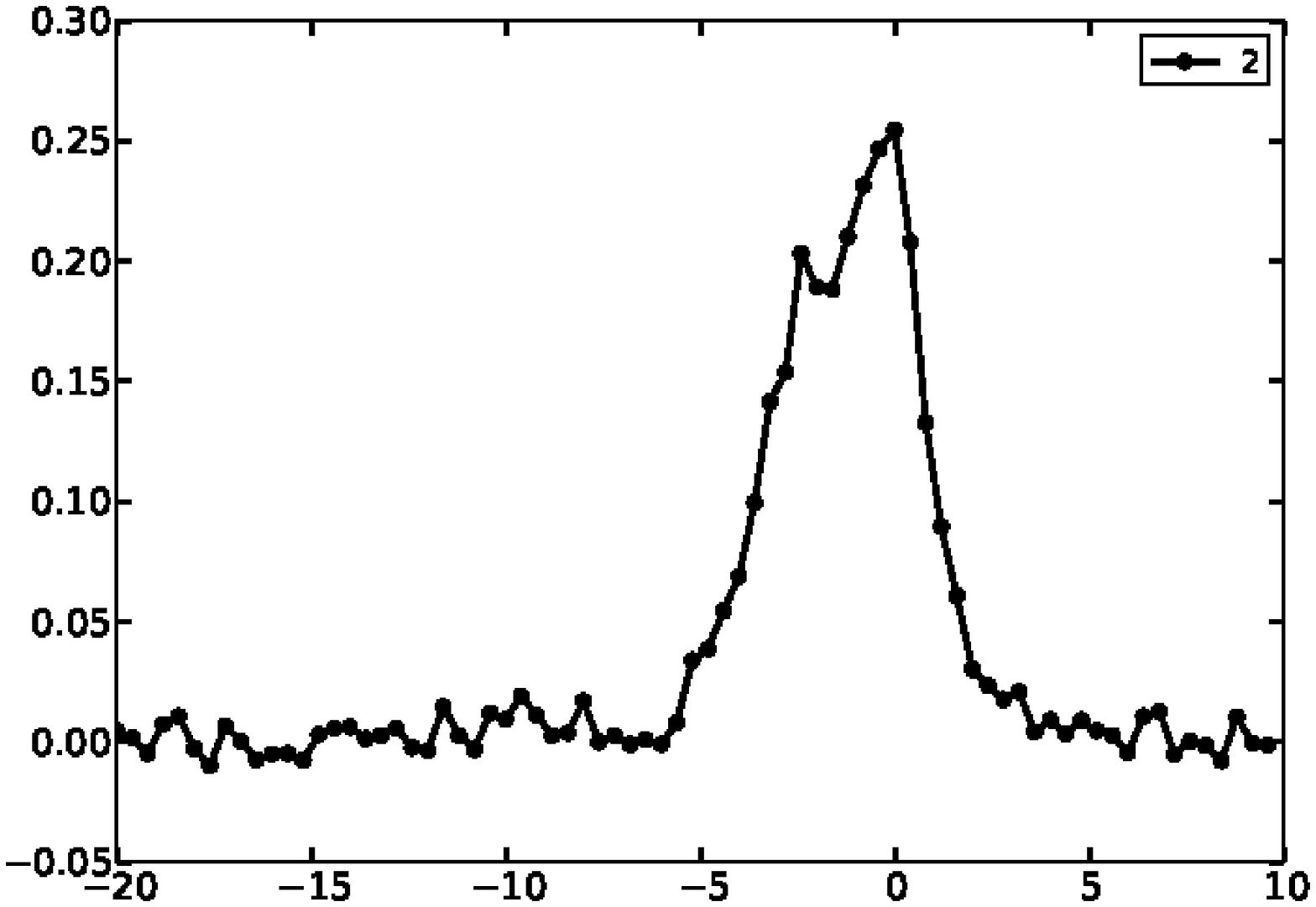}
\includegraphics[scale=0.30]{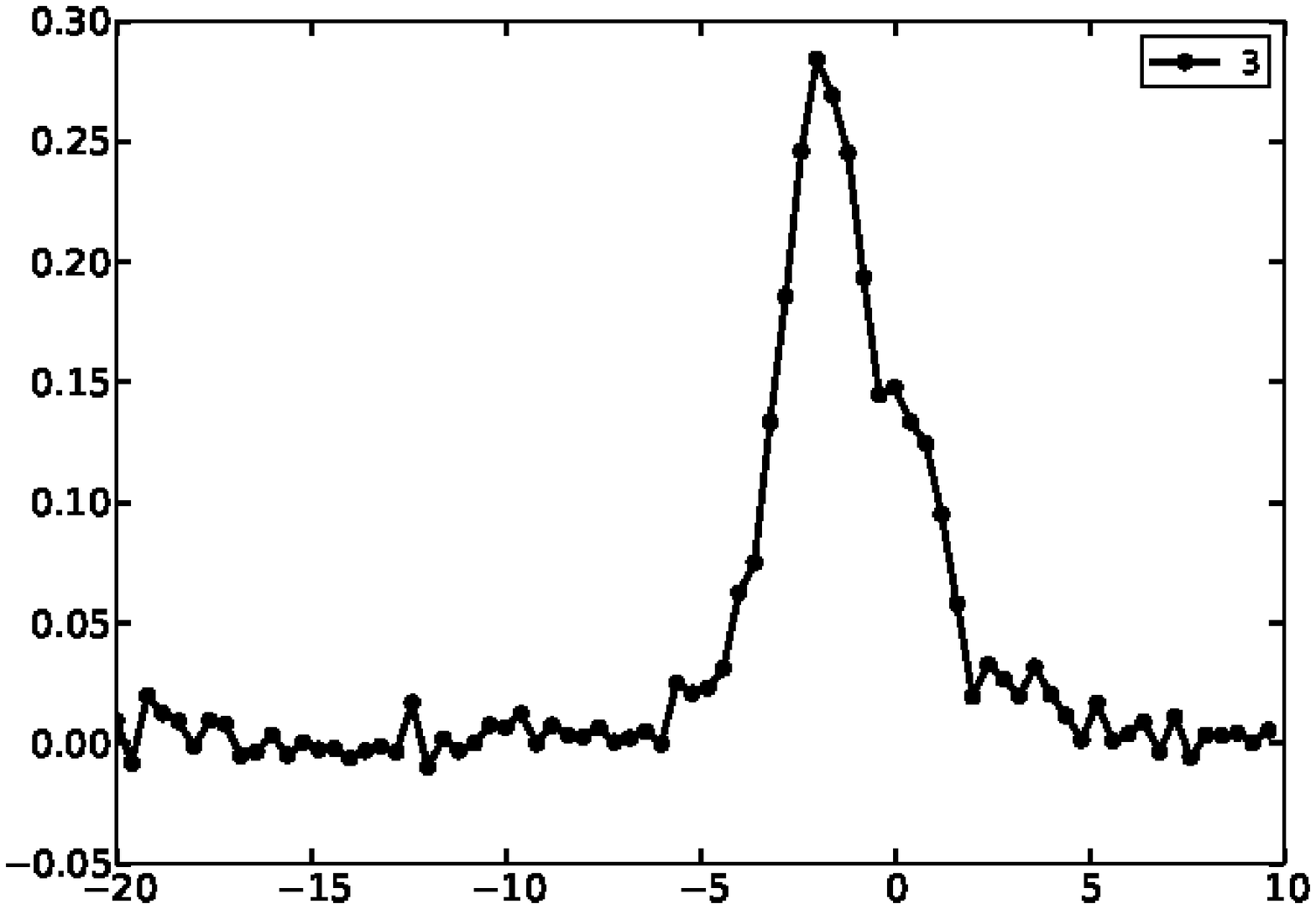}\\
\includegraphics[scale=0.30]{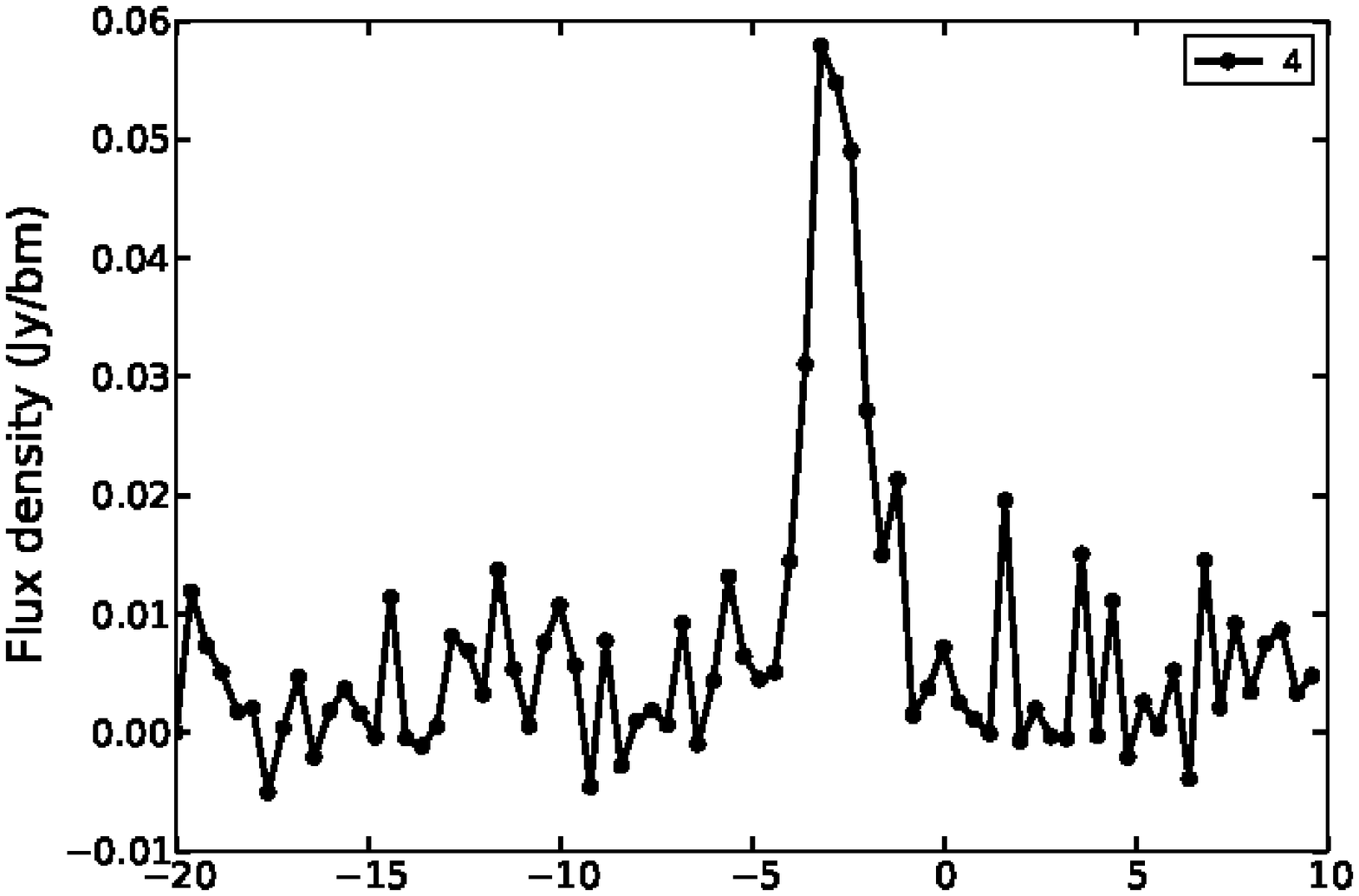}
\includegraphics[scale=0.30]{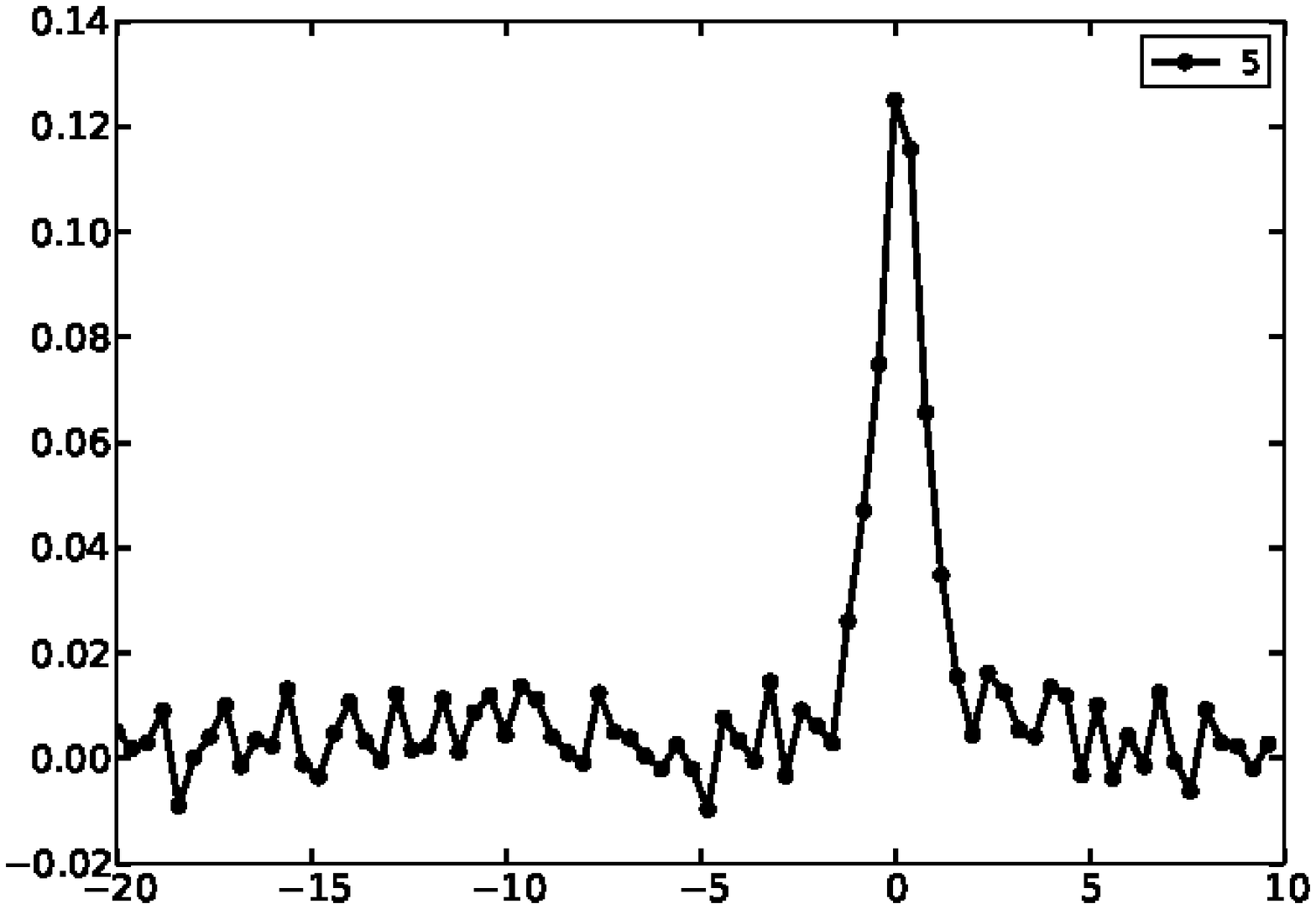}
\includegraphics[scale=0.30]{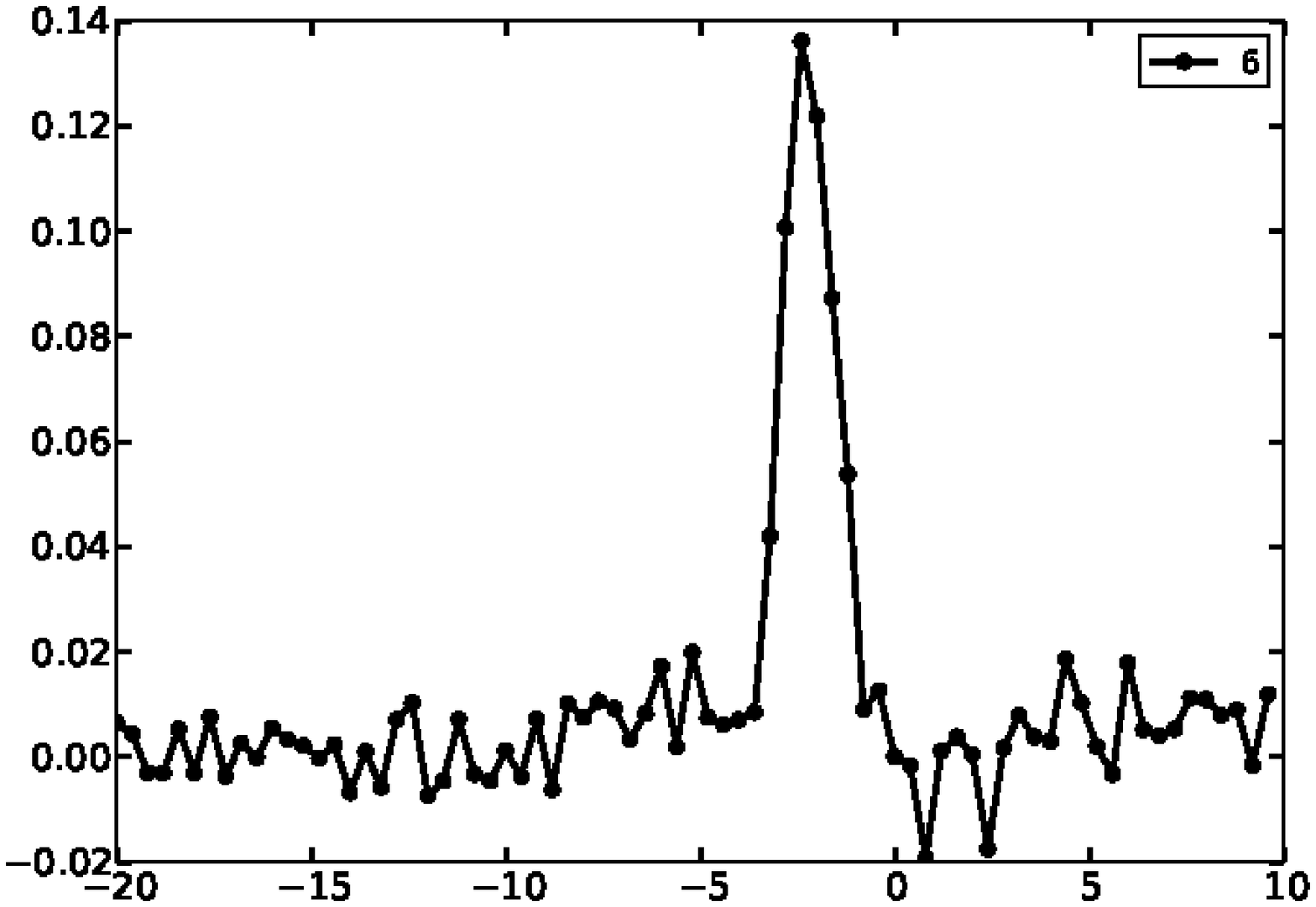}\\
\includegraphics[scale=0.30]{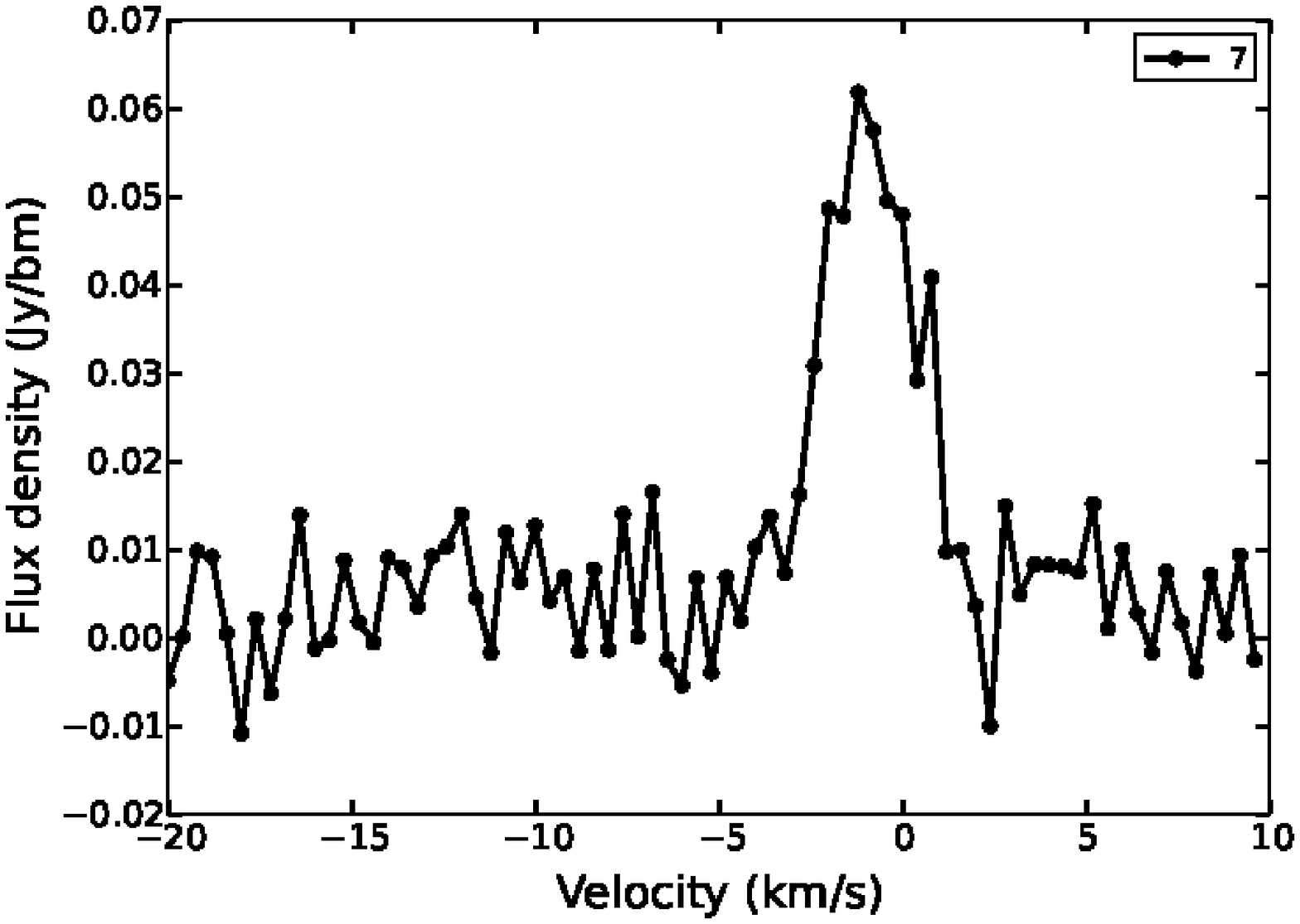}
\includegraphics[scale=0.30]{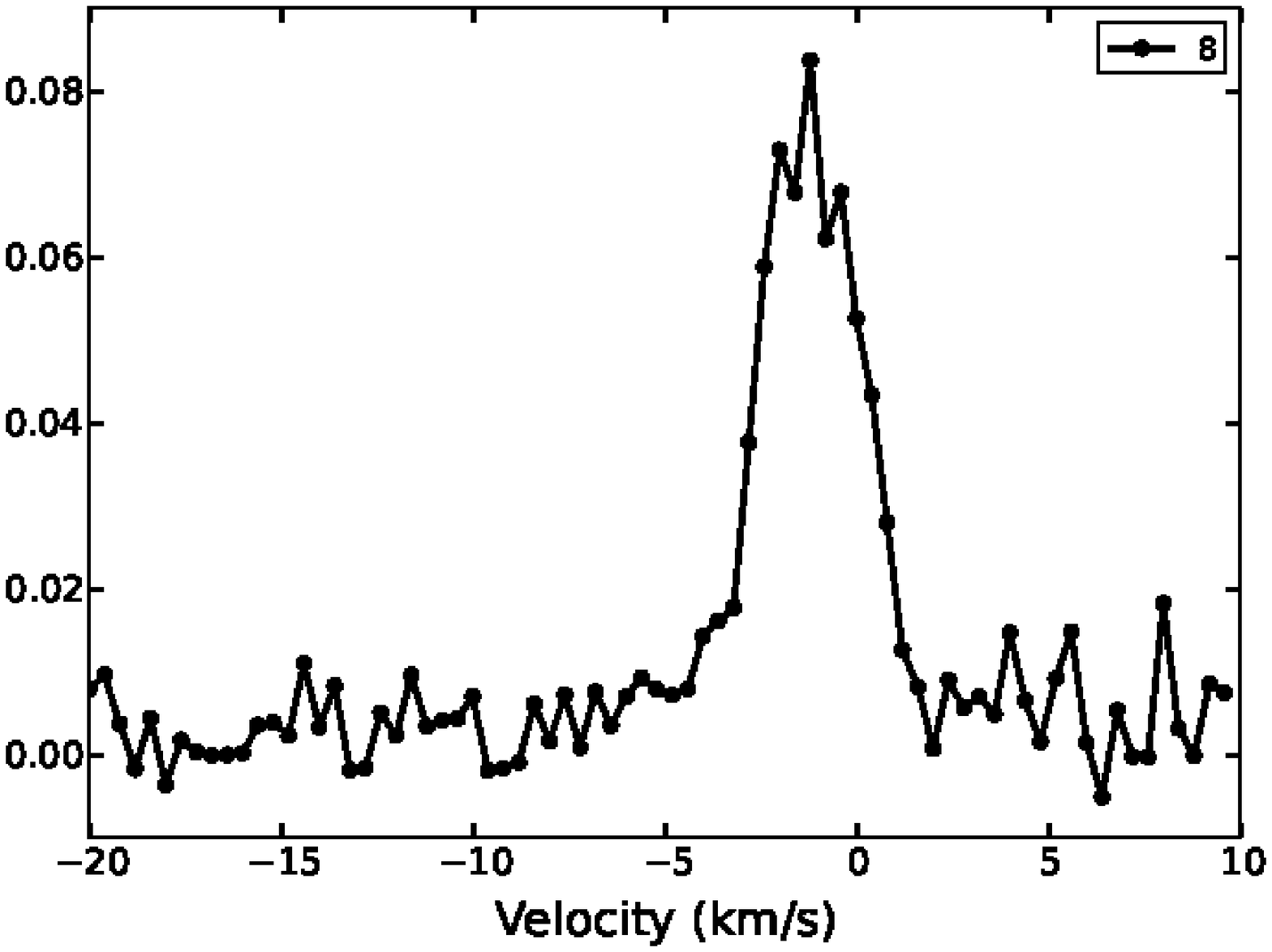} 
\includegraphics[scale=0.30]{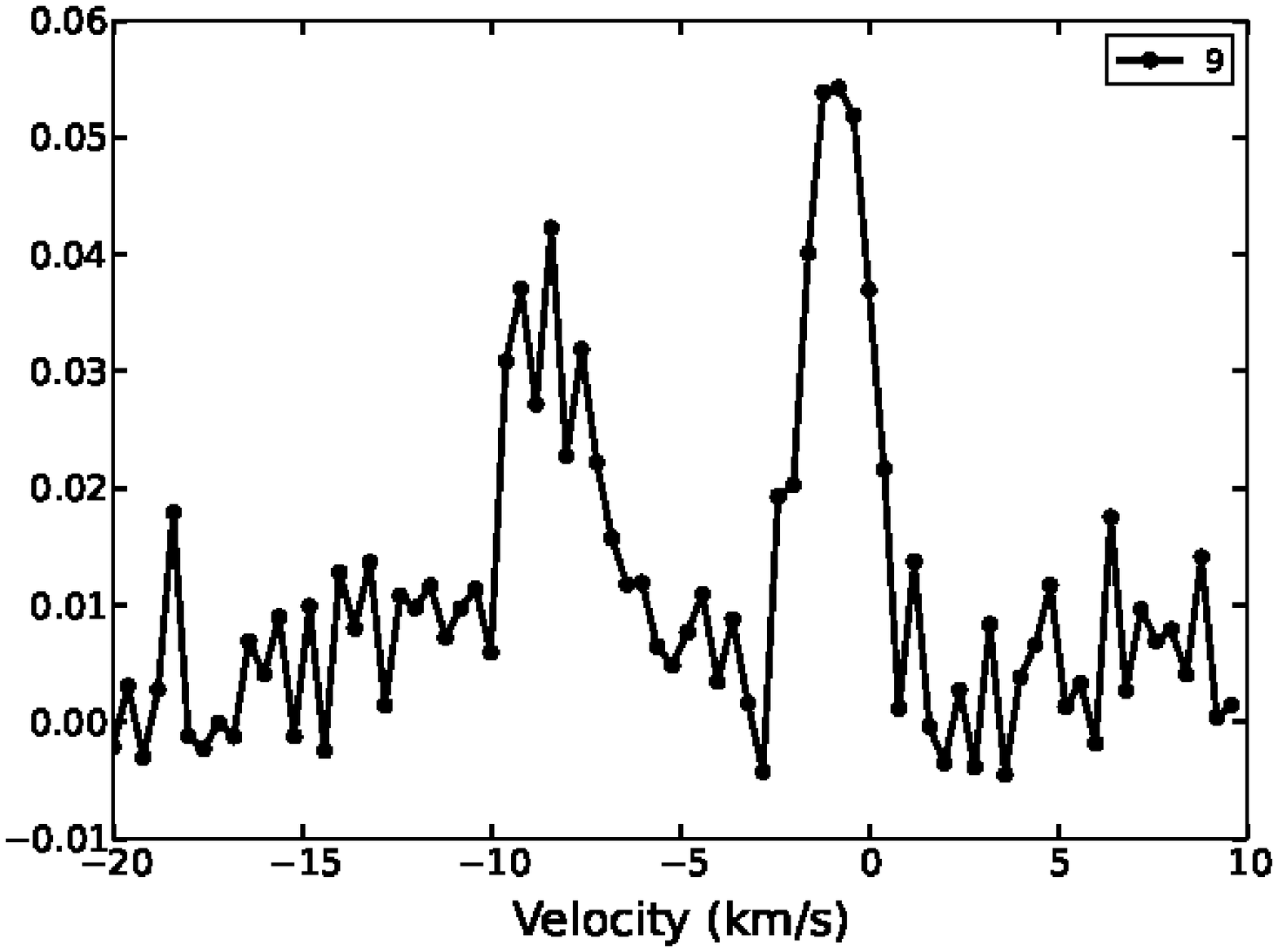}\\
\includegraphics[scale=0.30]{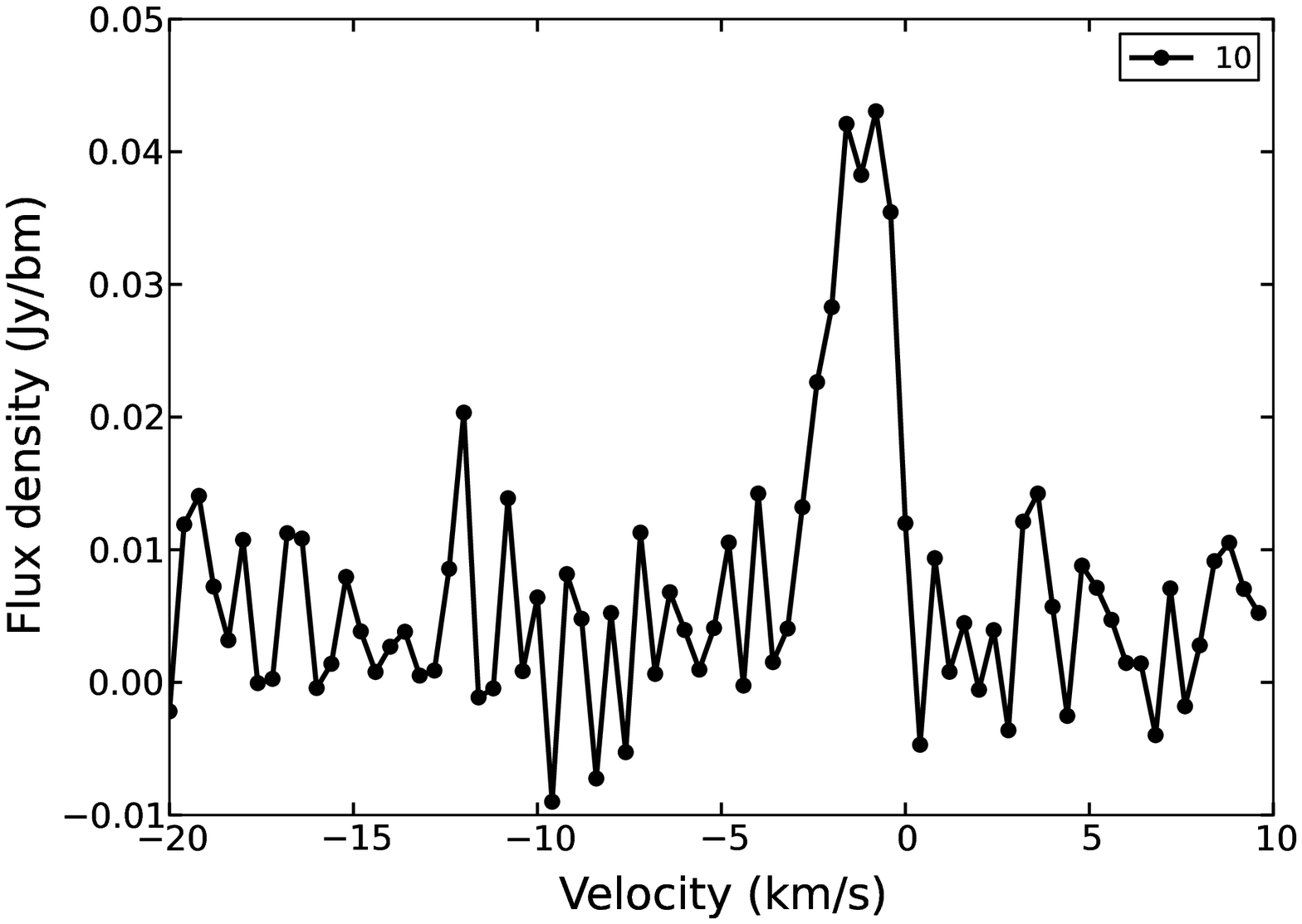}
\caption {\scriptsize {The HCO$^{+}$($1-0$) emission spectra of 10 molecular clumps of the N\,55-S. We note an additional red-shifted feature in the spectrum of clump 9, possibly due to any dynamical activity in this region.}}
\label{fig:spectra-hcop1}
\end{center}
\end{figure*}

We present HCO$^{+}$($1-0$) emission spectra of all ten identified clumps of the N\,55-S in Fig \ref{fig:spectra-hcop1}. We assert that all detections are strong with HCO$^{+}$(1$-$0) emission peaks detected with a minimum spectral signal to noise ratio of 4. We note an additional red-shifted feature in the spectrum of clump 9, possibly due to any dynamical activity in this region. The HCN/HCO$^{+}$ flux ratio of the clumps ranges from 0.46 $\pm$ 0.17 to 0.78 $\pm$ 0.12 (see Table \ref{tab:clump-details-astrodendro}), indicating overall weaker HCN flux in N\,55-S compared to HCO$^{+}$. We use HCO$^{+}$ as the primary clump tracer and further investigate various clump physical properties.

\subsection{Clump column density}
\label{sec:4.1}
The molecular column density of the HCO$^{+}$($1-0$) transition at frequency $\nu$ is obtained by the assumption of Local Thermodynamic Equilibrium \citep[LTE;][]{barnes2011,mangum2015}
\begin{equation}
\label{eqn:colundensity-first}
    N=\frac{3h}{8\,\pi^{3}\mu_{lu}^{2}} \frac{Q_{\rm rot}}{g_{\rm u}} \rm exp\left( \frac{E_{\rm u}}{kT_{\rm ex}}\right) \left[ \rm exp \left( \frac{h\nu}{kT_{\rm ex}}\right) - 1 \right] \int \tau_{\nu}dV
\end{equation}
Here $\mu_{lu}$ is the electric dipole matrix element which can be defined as $\mu_{lu}$ = $S\mu^{2}$, where $S$ is the line strength and $\mu$ is the dipole moment. $Q_{\rm rot}$ is the rotational partition function of the HCO$^{+}$ molecule,
\begin{equation}
    Q_{\rm rot} \equiv \Sigma_{i}g_{i}\, \rm exp \left( -\frac{E_{\rm i}}{kT} \right)
\end{equation}
where $g$ is the degeneracy of the corresponding rotational level. $E_{\rm u}$ and $g_{\rm u}$ denote the energy and degeneracy of the upper molecular level respectively. $T_{\rm ex}$ is the excitation temperature and $\int \tau_{\nu}\,dV$ denotes the optical depth of the emission line integrated over the velocity range. 

The radiative transfer equation in the Rayleigh-Jeans limit can be written as
\begin{equation}
    T_{\rm p} = (T_{\rm ex}-T_{\rm bg})(1-e^{-\tau})
\end{equation}
Here $T_{\rm p}$ is the peak brightness temperature of the emission line, and $T_{\rm bg}$ is the background temperature taken to be 2.72 K. The excitation temperature ($T_{\rm ex}$) can be precisely determined if we have multiple transitions of HCO$^{+}$, while our observation is limited to single transition. The excitation temperature $T_{\rm ex}$ of the cloud can also be estimated from optically thick $^{12}$CO ($1-0$) transition. \cite{naslim2018} determined the excitation temperature of the N\,55-S molecular cloud using the $^{12}$CO ($1-0$) transition \cite[see Fig 4 of ][]{naslim2018}. Their study shows that $T_{\rm ex}$ values of $^{12}$CO ($1-0$) range from 20 to 40 K in the N\,55-S. We assume $T_{\rm ex}=30$\,K as the excitation temperature of the HCO$^{+}$(1$-$0) clumps in the N\,55-S for further calculations. This value is consistent with the typical excitation temperature of molecular gas in massive clumps \citep{faundez2004,fontani2005}. We obtain the peak optical depth ($\tau_{\rm p}$) from the peak brightness temperature of each clump using the equation \citep{barnes2011}
\begin{equation}
\label{eqn:taup}
    \tau_{\rm p} = -ln\left[ 1-\frac{T_{\rm p}}{(T_{\rm ex}-T_{\rm bg})} \right]
\end{equation}
The peak brightness temperature $T_{\rm p}$ and optical depth $\tau_{\rm p}$ of each clump are tabulated in Table \ref{tab:flux ratio and mass surface density}. The partition function $Q_{rot}$ of HCO$^{+}$($1-0$) transition is 14.4 at $T_{\rm ex}=30$\,K \citep{rohlfs2004,mangum2015}. Substituting for $Q_{rot}$, and taking $\int \tau_{\nu}dV = \tau_{\rm p} \int \phi$ dV = $\tau_{\rm p} \sqrt{2\pi}\sigma_{\rm V}$, equation \ref{eqn:colundensity-first} can be simplified as \citep{barnes2011},
\begin{equation}
    N_{\rm p} = 6.02 \times 10^{17} \tau_{\rm p} \sqrt{2\pi} \sigma_{\rm V} \, \rm m^{-2}
\end{equation}
Here, we assume the emission line profiles, $\phi(V)$, to be Gaussian and $\sigma_{\rm V}$ denotes the velocity dispersion of the line in km\,s$^{-1}$. Assuming the relative abundance of HCO$^{+}$ to H$_{2}$ to be X=10$^{-9}$ \citep{garrod2008,loren1990,caselli2002,lee2003,zinchenko2009}, we derive the column density of molecular hydrogen ($N_{\rm H_{2}}$). The estimated values of $N_{\rm H_{2}}$ are tabulated in Table \ref{tab:flux ratio and mass surface density}. Clumps which are truly at a higher $T_{\rm ex}$ will have lower $N_{\rm p}$ and viceversa. For  $T_{\rm ex}$ = 40 K, the $N_{\rm p}$ values of the clumps are $\sim$ 28\% lower compared to the values derived for $T_{\rm ex}$ = 30\,K and $\sim$ 39\% higher for $T_{\rm ex}$ = 20\,K.

\subsection{Clump volume density}
The H$_{2}$ volume density ($n_{\rm col}$) of each clump can be calculated assuming that the physical depth of the source is comparable to its projected size \citep{barnes2011}. 
\begin{equation}
    n_{\rm col} = \sqrt{\frac{\rm ln 2}{\pi}} \frac{N_{\rm p}}{RX}
\end{equation}
The above equation provides an average volume density through the clump along the peak of the emission line. 
The derived clump volume densities range from 300 $\pm$ 50 to 4850$\pm$1630 cm$^{-3}$ (Table \ref{tab:flux ratio and mass surface density}). HCO$^{+}$($1-0$) is expected to be thermalized at a critical density of $n_{\rm cr} \sim 3 \times 10^{5}$ cm$^{-3}$ \citep{barnes1990}. The HCO$^{+}$($1-0$) line emission towards all the clumps in our sample is excited well below the critical density.

\subsection{Mass surface density}
We estimate the total mass surface density ($\Sigma_{\rm p}$) of molecular clumps using the HCO$^{+}$(1$-$0) column densities \citep{barnes2011}
\begin{equation}
\label{eqn:mass-surface-density}
    \Sigma_{\rm p} = \left( \frac{N_{\rm p}}{X} \right) (\mu_{\rm mol}m_{\rm H})
\end{equation}
where $\mu_{\rm mol}$ is the mean molecular weight in the gas which is taken to be 2.3 \citep{barnes2011}. The calculated mass-surface density of all identified clumps of N\,55-S are given in Table \ref{tab:flux ratio and mass surface density}. The values of mass-surface density vary from 29 $\pm$ 4 to 290 $\pm$ 41 $M_{\odot}$\,pc$^{-2}$.

\subsection{Mass of molecular clumps}
The masses of molecular clumps ($M_{\rm LTE}$) are calculated from the derived HCO$^{+}$($1-0$) column densities under the LTE assumption.
\begin{equation}
\label{eqn:LTE mass}
    M_{\rm LTE} = \frac{N_p}{X} (\mu_{\rm mol}m_{\rm H})\, \pi R^{2}
\end{equation}
We also determine the clump virial masses \citep{larson1981,solomon1987,saito2006,wong2006,muller2010} using the equation,
\begin{equation}
    M_{\rm vir} = 125 M_{\odot} \frac{(5-2\beta)}{(3-\beta)} \Delta v^{2} R
\end{equation}
This equation is based on the assumption that clumps are spherical in shape with a radial power-law density profile of index $\beta$. we assume $\beta = 1$ \citep{vandertak2000}. $R$ is the clump radius in parsec and $\Delta v$ is the FWHM of the emission line in km\,s$^{-1}$.   
The $M_{\rm LTE}$ and $M_{\rm vir}$ masses of the clumps, as well as the virial parameter, $\alpha$ = $M_{\rm vir}/M_{\rm LTE}$ , are tabulated in Table \ref{tab:flux ratio and mass surface density}.  

\begin{deluxetable*}{lccccccccc}
\tablecaption{Details of the clumps identified by astrodendro  \label{tab:clump-details-astrodendro}}
\tablehead{
\colhead{ID} & \colhead{RA} & \colhead{Dec} & \colhead{R} & \colhead{$\sigma_{\rm v}$} & \colhead{$F_{\rm HCN}$/} & \colhead{ YSO}\\
\colhead{No} & \colhead{(deg)} & \colhead{(deg)} &  \colhead{(pc)}  & \colhead{(km\,s$^{-1}$)} & \colhead{$F_{\rm HCO^{+}}$} & \colhead{association}
}
\startdata
1  &  83.1406  &  -66.4551  &   0.56$\pm$0.14  &  0.43$\pm$0.10    &  0.78$\pm$0.12   &  No    \\ 
2  &  83.1374  &  -66.4523  &   0.49$\pm$0.15  &  0.86$\pm$0.12    &  0.60$\pm$0.08   & Yes    \\ 
3  &  83.1343  &  -66.4541  &   0.48$\pm$0.19  &  0.67$\pm$0.13    &  0.77$\pm$0.15   & Yes    \\ 
4  &  83.1560  &  -66.4450  &   1.11$\pm$0.07  &  0.62$\pm$0.05    &  0.58$\pm$0.07   &  No    \\   
5  &  83.1385  &  -66.4617  &   0.68$\pm$0.12  &  0.53$\pm$0.09    &  0.51$\pm$0.09   &  No    \\ 
6  &  83.1244  &  -66.4557  &   0.41$\pm$0.19  &  0.48$\pm$0.19    &  0.46$\pm$0.17   &  No    \\ 
7  &  83.0933  &  -66.4601  &   0.79$\pm$0.10  &  0.94$\pm$0.07    &  0.54$\pm$0.12   & Yes    \\ 
8  &  83.0955  &  -66.4526  &   1.12$\pm$0.07  &  1.02$\pm$0.04    &  0.53$\pm$0.08   & Yes    \\ 
9  &  83.1347  &  -66.4592  &   0.85$\pm$0.11  &  0.62$\pm$0.07    &  --     & No     \\ 
10 &  83.0821  &  -66.4490  &   0.79$\pm$0.10  &  0.54$\pm$0.08    &  --     & No     \\
\enddata
\tablecomments{$R$ and $\sigma_{\rm v}$ denotes the radius and velocity dispersion of each clump derived from astrodendro. $F_{\rm HCO^{+}}$ and $F_{\rm HCN}$ are the flux densities of HCO$^{+}$(1$-$0) and HCN(1$-$0) clumps respectively.}
\end{deluxetable*}

\begin{deluxetable*}{lccccccccccccc}
\tablecaption{Properties of HCO$^{+}$(1$-$0) clumps.  \label{tab:flux ratio and mass surface density}}
\tablehead{
\colhead{ID}  & \colhead{$\Delta v$} & \colhead{$T_{\rm p}$} & \colhead{$\tau_{\rm p}$}  & \colhead{$M_{\rm vir}$} & \colhead{$N_{\rm H_{2}}$} & \colhead{$M_{\rm LTE}$} & \colhead{$\Sigma_{\rm p}$} & \colhead{$n_{col}$} & \colhead{$\alpha$}\\
\colhead{No}  &  \colhead{(km\,s$^{-1}$)} & \colhead{(K)} & -  & \colhead{($\times$10$^{2} {\rm M}_{\odot}$)} & \colhead{($\times$10$^{21}$cm$^{-2}$)} & \colhead{($\times$10$^{2} {\rm M}_{\odot}$)} & \colhead{($\times$10$^{2} {\rm M}_{\odot}$\,pc$^{-2}$)} & \colhead{($\times$10$^{3}$cm$^{-3}$)} & \colhead{($M_{\rm vir}$/$M_{\rm LTE}$)} 
}
\startdata
1  &  1.01$\pm$0.24  & 2.75  & 0.10  &  1.07 $\pm$0.57 &   6.85 $\pm$ 1.66 & 1.24 $\pm$ 0.67   &   1.26 $\pm$ 0.31   &  1.87 $\pm$ 0.64 &  0.86 $\pm$ 0.66 \\  
2  &  2.03$\pm$0.29  & 3.10  & 0.12  &  3.82 $\pm$1.59 &   15.71$\pm$ 2.22 & 2.22 $\pm$ 1.39   &   2.90 $\pm$ 0.41   &  4.85 $\pm$ 1.63 &  1.73 $\pm$ 1.30 \\  
3  &  1.58$\pm$0.30  & 3.46  & 0.13  &  2.26 $\pm$1.22 &   13.76$\pm$ 2.61 & 1.84 $\pm$ 1.46   &   2.54 $\pm$ 0.48   &  4.36 $\pm$ 1.87 &  1.23 $\pm$ 1.18 \\  
4  &  1.45$\pm$0.12  & 0.71  & 0.03  &  4.35 $\pm$0.79 &   2.43 $\pm$ 0.21 & 1.72 $\pm$ 0.27   &   0.45 $\pm$ 0.04   &  0.33 $\pm$ 0.03 &  2.53 $\pm$ 0.61 \\  
5  &  1.26$\pm$0.20  & 1.52  & 0.06  &  2.03 $\pm$0.74 &   4.62 $\pm$ 0.75 & 1.25 $\pm$ 0.48   &   0.85 $\pm$ 0.14   &  1.03 $\pm$ 0.24 &  1.62 $\pm$ 0.85 \\  
6  &  1.12$\pm$0.45  & 1.66  & 0.06  &  0.96 $\pm$0.90 &   4.51 $\pm$ 1.82 & 0.43 $\pm$ 0.45   &   0.83 $\pm$ 0.33   &  1.69 $\pm$ 1.05 &  2.22 $\pm$ 3.11 \\  
7  &  2.20$\pm$0.16  & 0.75  & 0.03  &  7.20 $\pm$1.39 &   3.95 $\pm$ 0.28 & 1.43 $\pm$ 0.38   &   0.73 $\pm$ 0.05   &  0.76 $\pm$ 0.11 &  5.04 $\pm$ 1.66 \\  
8  &  2.40$\pm$0.10  & 1.02  & 0.04  &  12.16$\pm$1.29 &   5.86 $\pm$ 0.24 & 4.29 $\pm$ 0.59   &   1.08 $\pm$ 0.04   &  0.79 $\pm$ 0.06 &  2.84 $\pm$ 0.49 \\   
9  &  1.46$\pm$0.17  & 0.66  & 0.02  &  3.43 $\pm$0.92 &   2.30 $\pm$ 0.27 & 0.97 $\pm$ 0.27   &   0.42 $\pm$ 0.05   &  0.41 $\pm$ 0.07 &  3.53 $\pm$ 1.37 \\  
10 &  1.26$\pm$0.19  & 0.52  & 0.02  &  2.36 $\pm$0.76 &   1.57 $\pm$ 0.23 & 0.56 $\pm$ 0.16   &   0.29 $\pm$ 0.04   &  0.30 $\pm$ 0.05 &  4.19 $\pm$ 1.82 \\    
\enddata
\tablecomments{$\Delta v$ denotes the FWHM of the HCO$^{+}$(1-0) emission line. $T_{\rm p}$ is the peak brightness temperature and $\tau_{\rm p}$ is the corresponding optical depth. $M_{\rm vir}$ denotes the virial masses of each clump. $N_{\rm H_{2}}$ denotes the column density of molecular hydrogen in each clump which is related to HCO$^{+}$ column density ($N_{\rm p}$) as $N_{\rm H_{2}}$ = $N_{\rm p}$ $\times$ 10$^{9}$. $M_{\rm LTE}$ and $\Sigma_{\rm p}$ denotes the LTE masses and the total mass surface density of each clump respectively. $n_{\rm col}$ denotes volume number density of molecular hydrogen in each clump and $\alpha$ represents the virial ratio defined as $M_{\rm vir}$/$M_{\rm LTE}$. The values are estimated assuming $T_{\rm ex}$ = 30 K. The uncertainties due to the assumption of $T_{\rm ex}$ described in \ref{sec:4.1} are not included in the errors here.}
\end{deluxetable*}

\section{Physical properties of dense molecular gas}
\label{sec:physical properties of dense molecular gas}
We investigate the physical properties of the dense molecular clumps of  N\,55-S in the LMC, traced by ALMA observations of HCO$^+$($1-0$) and HCN($1-0$) emission. 
In Fig \ref{fig:all-clumps}, we compare the velocity integrated intensity maps of $^{12}$CO($1-0$) and $^{13}$CO($1-0$) \citep{naslim2018} with HCO$^+$($1-0$) and HCN($1-0$) of the N\,55-S region. The spatial distribution of the emission from all four molecular transitions are broadly similar. The maximum intensity in all four emission maps comes from clumps 2 and 3. The $^{13}$CO($1-0$), HCO$^{+}$($1-0$) and HCN($1-0$) emission are not detected in regions of weak $^{12}$CO($1-0$) emission. It is evident that $^{12}$CO($1-0$) emission is more widely spread compared to $^{13}$CO($1-0$), HCO$^{+}(1-0)$ and HCN($1-0$). The distribution of the HCO$^{+}(1-0)$ and HCN($1-0$) emission is most compact. The projected spatial distribution of emission from multiple species reflects the density structure of the molecular clumps from surface to the interior.  

\subsection{Dense gas fraction}
\label{subsec:dense gas fraction}
We estimate the H$_{2}$ mass traced by $^{12}$CO($1-0$) \citep{naslim2018} luminosities ($L_{\rm CO}$) from the N\,55-S region using the equation \citep{wong2011}.
\begin{equation}
M_{\rm H_{2}} [M_{\odot}] = 4.4 \frac{X_{\rm CO}}{2.2 \times 10^{20} \, \rm cm^{-2} (\rm K\,km\,s^{-1})} L_{\rm CO} (\rm K\,km\,s^{-1}\,pc^{2})
\end{equation}
 Here $X_{\rm CO}$ denotes the CO-to-H$_{2}$ conversion factor which has a Galactic value, $X_{\rm CO}$=2$\times$10$^{20}$ cm$^{-2}$ K$^{-1}$ km$^{-1}$ s \citep{strong1988, bolatto13}. Both theoretical and observational studies suggest that $X_{\rm CO}$ value increases with decreasing metallicity. However, a Galactic value can be approximated for environments with metallicities down to $\sim$0.5\,$Z_{\odot}$ \citep{leroy2011, bolatto13, pineda2017}. Hence we use $X_{\rm CO}$=2$\times$10$^{20}$ cm$^{-2}$ K$^{-1}$ km$^{-1}$ s in our calculation. \cite{naslim2018} report a 30\% missing flux in the ALMA $^{12}$CO(1-0) emission maps. Accounting for this, 
the total H$_{2}$ mass traced by $^{12}$CO(1$-$0) is (2.59 $\pm$ 0.01) $\times$ 10$^{4}$ $M_{\odot}$. The total H$_{2}$ mass traced by HCO$^{+}$(1$-$0) is $\sim$ (0.70 $\pm$ 0.10) $\times$ 10$^{3}$ $M_{\odot}$ (sum of $M_{\rm LTE}$/$\mu_{\rm mol}$ in Table \ref{tab:flux ratio and mass surface density}). Thus the dense gas fraction in the N\,55-S region is 0.025$\pm$0.005. A higher value of  $X_{\rm CO}$ = 4$\times$10$^{20}$ cm$^{-2}$ K$^{-1}$ km$^{-1}$ s \citep{hughes2010} will further decrease the dense gas fraction to 0.013.

\subsection{Low volume densities of dense gas clumps}
\label{subsec:low volume densities of dense gas clumps}
The volume densities of ten identified clumps are in a range ($0.30-4.85$)$\times 10^3$ cm$^{-3}$ (Table \ref{tab:flux ratio and mass surface density}). Since HCO$^+(1-0)$ has a relatively high critical density $\sim 3 \times 10^5$ cm$^{-3}$ \citep{barnes1990}, we would expect the bulk of HCO$^{+}(1-0)$ luminosity to originate in thermalized star-forming cores. However, all ten HCO$^+$ clumps in our samples show volume densities well below the expected critical density. We note that similar low volume densities are reported for Milky Way clouds from HCO$^+(1-0)$ emission \citep{barnes2011}. The authors report that 95\% of all massive clumps emit well below the critical density. A study of HCO$^+(1-0)$  emission from multiple clouds of the LMC also suggests that the majority of the clumps have volume densities\footnote{We computed the volume densities from the $M_{LTE}$ and $R$ values from Table 2 of \citep{seale2012}.} well below the critical density of $J=1-0$ line \citep{seale2012}.

This could happen if HCO$^{+}(1-0)$ is sub-thermally excited and not thermalized to the local H$_{2}$ gas. Significant HCO$^{+}(1-0)$ emission can arise at densities lower than the critical density of the line \citep{evans1999,shirley2015,kauffmann2017}. Another possibility is a severe underestimation of optical depth of the emission line due to small beam filling factor. The clump mass can be calculated assuming that the clump volume contains gas at the critical density of the HCO$^{+}$($1-0$) transition \citep{barnes2011}.
\begin{equation}
    M=5.3 \times f \left( \frac{n_{\rm cr} }{10^{11} \rm m^{-3} } \rm \right)\left( \frac{\rm R}{\rm pc}\right) ^{3}
\end{equation}
Here $n_{\rm cr}$ denotes the critical density of HCO$^{+}(1-0)$ and $f$ denotes the beam filling factor. Comparing the above derived mass with the cloud mass derived from LTE analysis yields $f=(0.001-0.01)$ for the clumps in our sample. This low beam filling factor could result from a highly clumpy structure of molecular clouds or if the clumps are not well resolved. 
\subsection{$M_{\rm vir}$ versus $M_{\rm LTE}$}
The virial and LTE masses of N\,55-S clumps are (0.96$-$12.16)$\times$10$^{2}$ M$_{\odot}$ and (0.43$-$4.29)$\times$10$^{2}$ $M_{\odot}$ respectively (see Fig \ref{fig:comparison-plot-seale2012}A). In order to examine whether these clumps are gravitationally bound, we inspect the virial parameter; $\alpha$=$M_{\rm vir}$/$M_{\rm LTE}$ (Fig \ref{fig:comparison-plot-seale2012}B). The average value of virial parameter is 2.6$\pm$1.2. A few isolated clumps such as 4, 6, 7, 8, 9 and 10 are not likely to be gravitationally bound.
%However, the virial parameters of these clumps are consistent with the gravitational equilibrium line within 2$\sigma$ error bar}. 
The relatively high value of virial parameter could be due to an underestimation of $M_{\rm LTE}$ in these region or due to dynamical impact of star formation. 

\subsection{size-linewidth relation}
\label{subsec:size-linewidth scaling}
\begin{figure*}
\begin{center}
\hspace{0.4cm}
\includegraphics[scale=0.4]{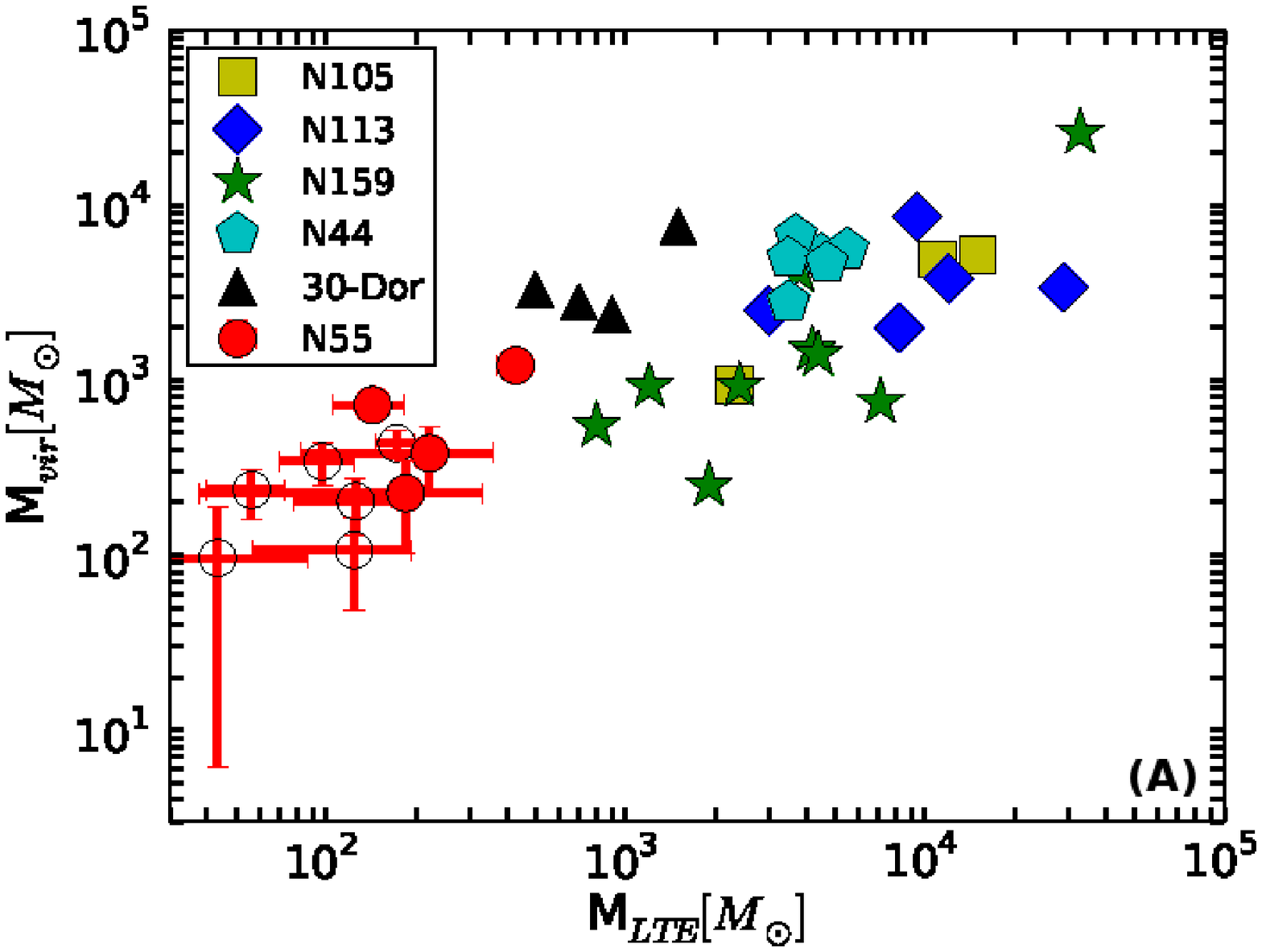}
\includegraphics[scale=0.4]{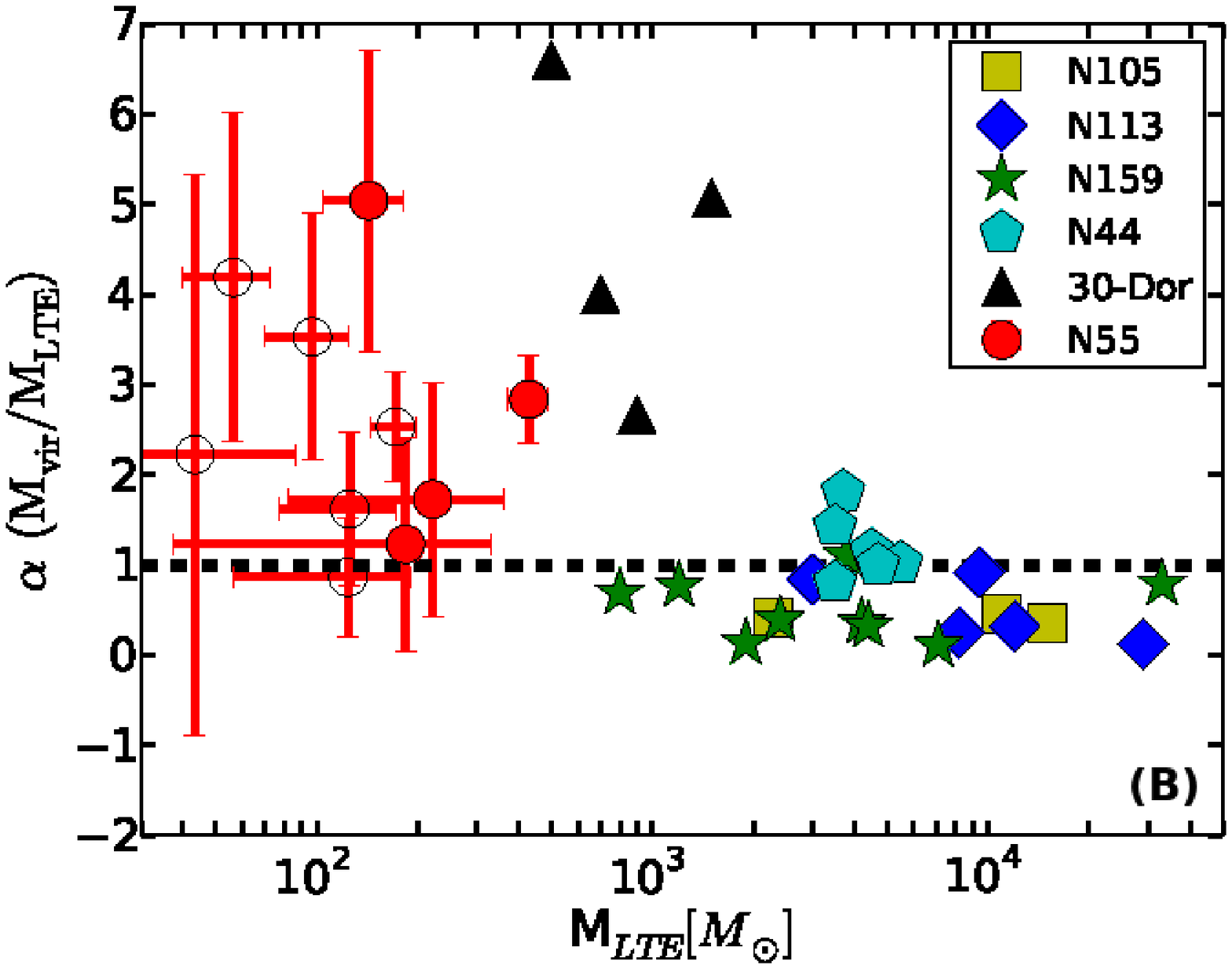}
\includegraphics[scale=0.4]{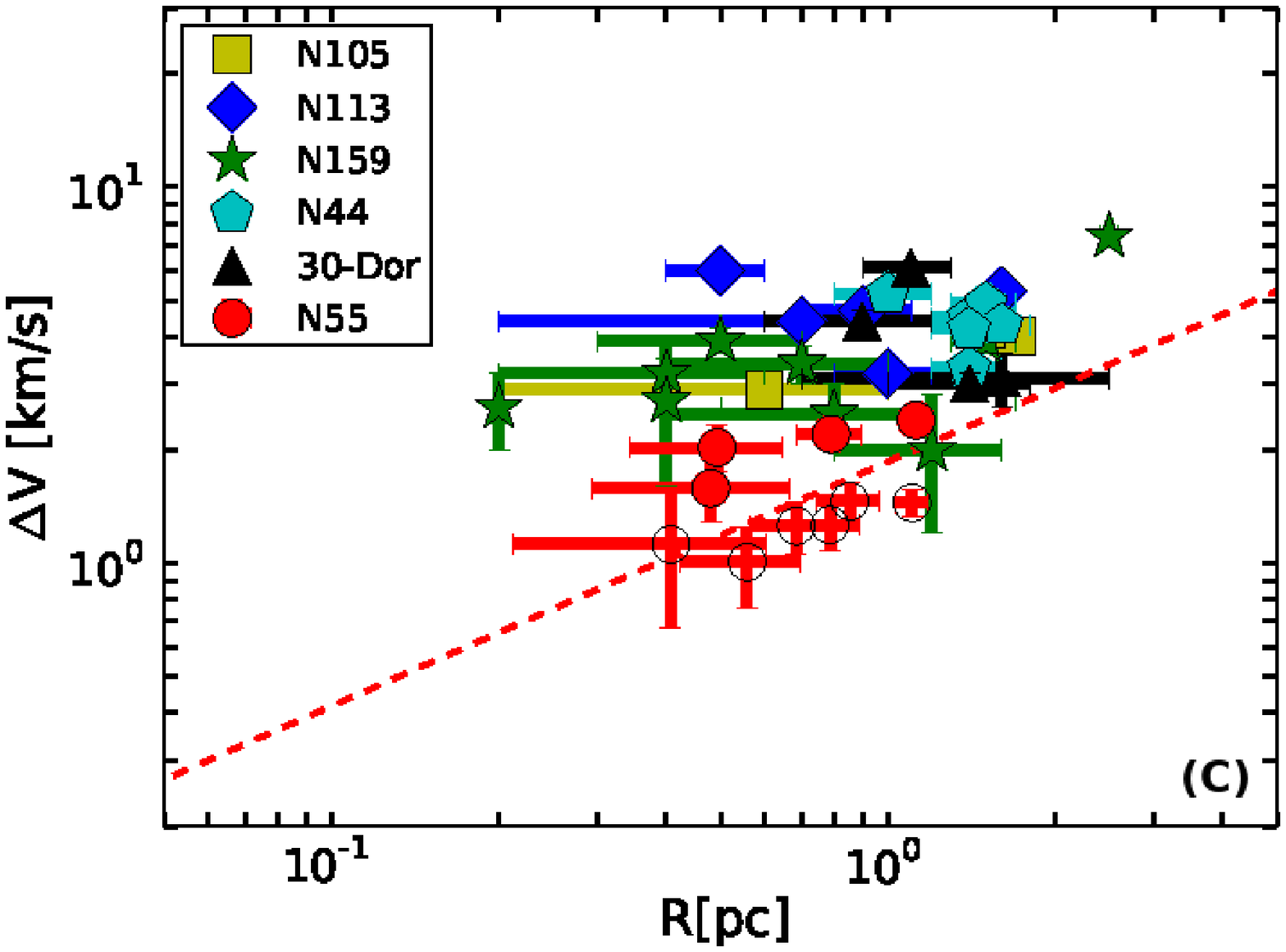}
\includegraphics[scale=0.38]{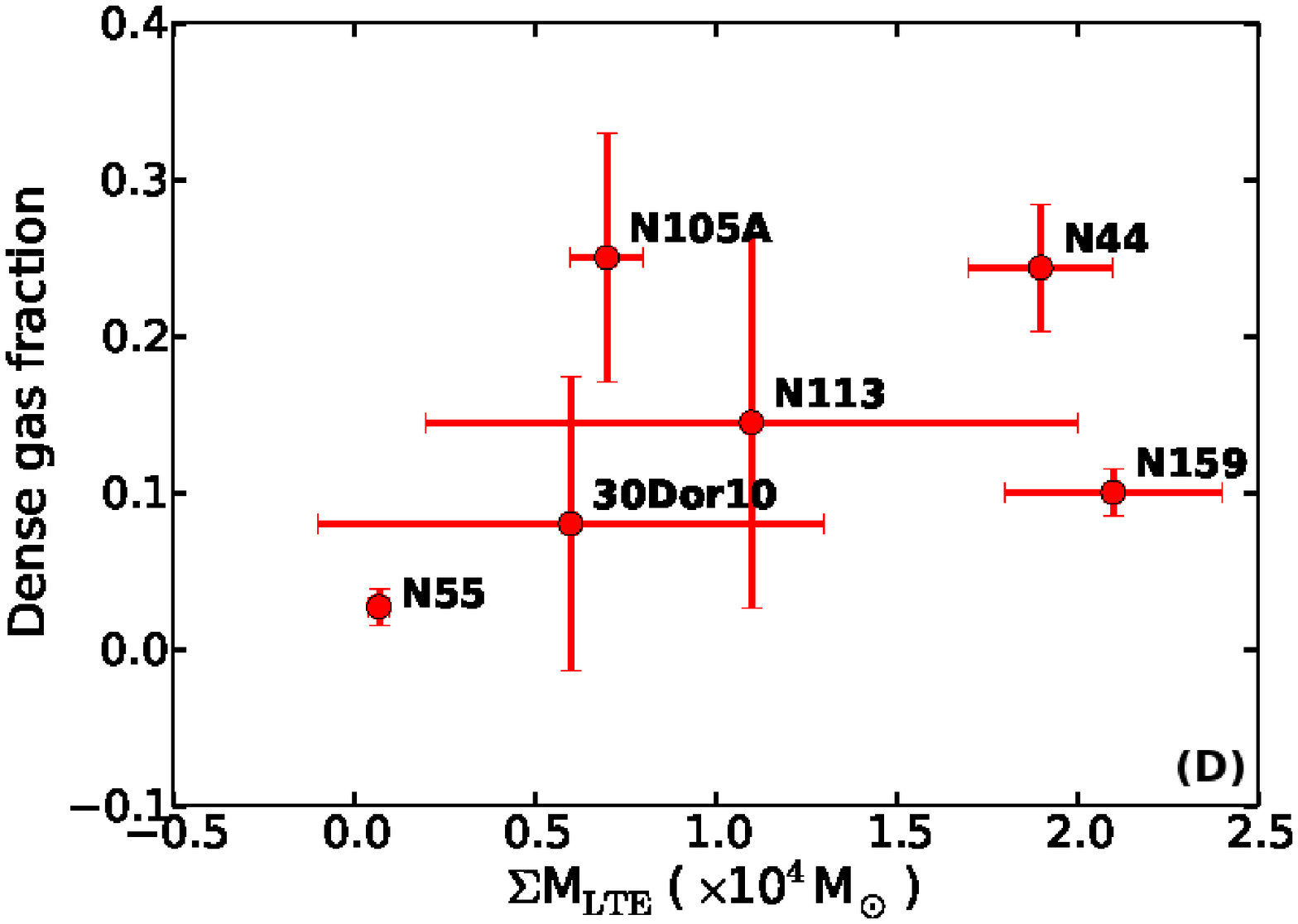}
\caption {\scriptsize (A) Virial versus LTE masses plot of HCO$^{+}$($1-0$) molecular clumps in N\,55-S (this work) along with clumps in other LMC GMCs N\,105, N\,113, N\,159, N\,44 \citep{seale2012} and 30\,Doradus-10 \citep{anderson2014}. (B) $M_{\rm LTE}$ versus virial parameter ($\alpha$) for the N\,55-S HCO$^{+}$($1-0$) clumps together with other GMCs of the LMC \citep{seale2012, anderson2014}. (C) The size versus linewidth relation for HCO$^{+}$($1-0$) clumps (red circles) in N\,55-S clumps fitted with a power-law, $\Delta v$ $\propto$ R$^{0.65 \pm 0.32}$ (red dotted line). The R and $\Delta v$ values of HCO$^{+}$($1-0$) clumps in other GMCs of the LMC  \citep{seale2012, anderson2014} are shown for comparison.  The closed cicles denote the clumps with associated YSOs and the open circles denote those without YSOs in panels A, B, and C. (D) The total molecular gas mass traced by HCO$^{+}$(1$-$0) under the assumption of LTE versus the dense gas mass fraction for N\,55-S and other GMCs of the LMC \citep{seale2012, anderson2014}.}
\label{fig:comparison-plot-seale2012}
\end{center}
\end{figure*}

The size and linewidth of HCO$^{+}$($1-0$) clumps follow a power-law, $\Delta v$ $\propto$ $R^{0.65\pm0.32}$ (Fig \ref{fig:comparison-plot-seale2012}C). The size-linewidth relation of $^{12}$CO($1-0$) clumps of the N\,55 is $\Delta v$ $\propto$ $R^{0.4}$ \citep{naslim2018}. The size-linewidth power-law index is found to be in the range $0.46-0.78$ for several dense gas tracers in the central molecular zone (CMZ) of the Milky Way \citep{shetty2012}. The slope is found to be 0.6 in extragalactic clouds \citep{bolatto2008} and 0.5 in Milky Way clouds \citep{heyer2009}. Thus the size-linewidth power-law index of HCO$^{+}$($1-0$) clumps of the N\,55-S is consistent with the CMZ, extragalactic, and the Milky Way clouds.

\section{Clump association with YSOs}
\label{sec:Clump association with YSOs}
There exist extensive catalogues of YSOs in the LMC due to its proximity and the ability to observe the entire galaxy \citep{whitney2008,gruendl2009,indebetouw2004,chen2009,chen2010,ellingsen2010}. This gives the advantage of positional matching the YSOs with dense molecular clumps to study star formation and the associated dense gas. In Fig \ref{fig:all-clumps}, we mark the position of YSOs in the N\,55-S \citep{gruendl2009,chen2009} on top of the velocity integrated intensity maps of HCO$^{+}(1-0)$ and HCN($1-0$). All 4 identified YSOs in the N\,55-S are near the HCO$^{+}(1-0)$ and HCN($1-0$) emission peaks (near clumps 2, 3, 7 and 8). The positional offset between the emission peak and YSO position is 1.16, 1.21, 0.43 and 1.69 arcsec for clump 2, 3, 7, and 8 respectively. If the YSOs are randomly distributed, the probability of a YSO locating within 2 arcsec of an emission peak is only 0.04\% . However, all 4 of the identified YSOs are near the emission peaks indicating their positions are not from a random distribution and are related to the dense molecular clumps. The young stars form in the densest cores of the molecular clouds \citep{krumholz2010}. The YSOs that are at significant offset from the emission peak of the molecular clumps are expected to be slightly evolved than those at the core. This hypothesis is observationally supported by the detection of maser emission (sign of early phase of YSO evolution) in 80$\%$ of YSOs located close to the emission peaks of molecular clouds \citep{seale2012}. Thus the YSOs associated  to clump 2, 3, 7, and 8 in the N\,55-S are likely to be early in their evolutionary stage. 

\cite{gruendl2009} suggest that the 8$\mu$m magnitude of YSOs can be treated as a good proxy of the YSO mass based on the radiation transfer models on the spectral energy distribution (SED) of various YSOs. The YSOs with 8$\mu$m magnitude brighter than [8.0] are classified as massive YSOs. The infrared SED fitting of various YSOs in the N\,44, and N\,159 shows YSO mass of 8$-$15 M$_{\odot}$ for 8$\mu$m magnitudes $\sim$ [8.76], [8.94], [8.24]. The masses are $\sim$ 5-10 M$_{\odot}$ for 8$\mu$m magnitude $\sim$ [10.17] \citep[Table 7 of ][]{chen2009}. The 8$\mu$m magnitudes of the YSOs associated with clumps 2, 3, 7, and 8 in the N\,55-S are $\sim$ [8.24], [7.27], [10.19] and [8.76] respectively \citep{gruendl2009}. Thus the YSOs associated to clump 2, 3, and 8 are likely to be massive (8$-$15 M$_{\odot}$) and associated to clump 7 could be of mass $\sim$ (5-10) M$_{\odot}$. 

We compare the physical properties of the clumps in light of the presence/absence of YSOs. The molecular clumps with YSOs in our sample are relatively massive than those without YSOs. The average mass of the YSO-associated clumps is 245 $\pm$ 96 $M_{\odot}$ whereas the average mass of clumps without YSOs is $\sim$ 103 $\pm$ 39 $M_{\odot}$. Similar lines of observational evidence are reported for various regions of the LMC \citep{seale2012,naslim2018} and for the Milky Way molecular clumps \citep{hill2005}.  The clumps with mass-surface densities $\geq$ 73 $M_{\odot}$\,pc$^{-2}$ shows YSO association in N\,55-S. The mass surface density threshold of clumps for massive star formation is $\sim$ 501 M$_{\odot}$\,pc$^{-2}$ and 794 $M_{\odot}$\,pc$^{-2}$ for N\,159W and N\,159E respectively \citep{nayak2018}. The star formation threshold for 30-Doradus is $\sim$ 670 $M_{\odot}$\,pc$^{-2}$ \citep{nayak2016}. Thus the mass density threshold of N\,159 and 30-Doradus is (6-10) times higher compared to N\,55-S. This could be due to the relatively less extreme star-forming environment of N\,55-S. The stronger radiation field in the N\,159 and 30-Doradus is possibly preventing massive star formation at low mass surface density in these regions.

We also find the velocity widths of YSO-bearing clumps to be systematically larger than those of non-YSO associated clumps. The velocity widths of YSO associated clumps are $\Delta v$ = $1.6-2.4$ km\,s$^{-1}$ whereas the clumps without YSOs, $\Delta v$ = $1.0-1.5$ km\,s$^{-1}$. The larger linewidths could be indicative of turbulence due to YSO activity, indicating that YSOs affect the properties of dense molecular clumps. These results are consistent with the $^{13}$CO($1-0$) and $^{12}$CO($1-0$) study of the N\,55 region \citep{naslim2018}. Similar results are reported for $^{12}$CO($2-1$) study of 30\,Doradus \citep{nayak2016}, 
$^{12}$CO/$^{13}$CO observations of N\,159W/E \citep{fukui2019,tokuda2019} and  HCO$^{+}$(1$-$0) study of several GMCs of the LMC \citep{seale2012}. 

\section{Comparison of N\,55 clump properties with other LMC clouds}
\label{sec:comparison with other LMC clumps}
\cite{seale2012} studied molecular clumps of different star forming GMCs of the LMC (N\,159, N\,105A, N\,44, and N\,113) using dense gas tracers HCO$^{+}$(1$-$0) and HCN(1$-$0). The authors used ATCA observations with a spatial resolution of $\sim$ 6$\times$7 arcsec$^{2}$ ($\sim$ 1.4$\times$1.7 pc$^{2}$ in linear scale) and a spectral resolution of $\sim$ 0.2 km\,s$^{-1}$ (for N\,105 and N\,113); 0.4 km\,s$^{-1}$ (N\,159 and N\,44). A similar study of dense gas tracers at same spatial resolution was carried out by \cite{anderson2014} in the 30\,Doradus-10 with a spectral resolution of 0.84 km\,s$^{-1}$. We examine the molecular clump properties such as size, linewidth, mass and dense gas fraction of N\,55-S region in comparison to other GMCs of the LMC in Fig \ref{fig:comparison-plot-seale2012}. 

The size of clumps in N\,55-S is comparable to the clumps of other GMCs (see Fig \ref{fig:comparison-plot-seale2012}C). However, the linewidths of the N\,55-S clumps are slightly small compared to other regions of the LMC \citep{anderson2014,seale2012}. The small linewidths could be indicative of relatively less energetic environmental conditions in the N\,55-S. 
We compare the clump masses (both virial and LTE) of N\,55-S with other star forming regions of the LMC \citep{seale2012} in Fig \ref{fig:comparison-plot-seale2012}A. The $M_{\rm vir}$ and $M_{\rm LTE}$ derived for N\,55-S clumps are systematically lower than other regions of the LMC. A few of the dense molecular clumps of N\,55-S are likely not in gravitational equilibrium. %However we note that the error bars on alpha values are high such that the alpha values are consistent with the gravitational equilibrium line within 3$\sigma$.
%mostly in gravitational equilibrium with a virial parameter $M_{\rm vir}$/$M_{\rm LTE}$ $\sim$ 1 (see Fig \ref{fig:comparison-plot-seale2012}B). 
%The $\alpha$ values of N\,55-S is similar to other GMCs except 30\,Doradus-10 which has relatively large values of virial parameter \citep{anderson2014}}.  

The molecular gas mass traced by the HCO$^{+}$(1$-$0) clumps versus dense gas fraction for N\,55-S and other GMCs \citep{seale2012,anderson2014} in the LMC are shown in Fig \ref{fig:comparison-plot-seale2012}D. The dense gas fraction of N\,55-S is 0.025$\pm$0.005, smaller than the fraction seen in other GMCs; 0.1$-$0.24. Dense gas fraction is crucial for massive star formation and is directly proportional to the star formation efficiency within a GMC \citep{krumholz2012,lada2012}. This suggests that N\,55-S has lower star formation efficiencies compared to other GMCs of the LMC. However, we note that different dynamical environments of galaxies and stellar feedback play crucial role in setting massive star formation efficiency \citep{querejeta2019,ochsendorf2017}.

The HCN/HCO$^{+}$ flux ratio of molecular clumps in the N\,55-S range from 0.46 $\pm$ 0.17 to 0.78 $\pm$ 0.12. HCN and HCO$^{+}$ molecules possess similar rotational constants and electric dipole moments. Hence the higher flux ratio can be attributed to the relative abundance. However we note that a strong UV radiation field due to high star formation activity can enhance the HCO$^{+}$ abundance in active star forming regions \citep{meijerink2011, bayet2011}.The HCN/HCO$^{+}$ flux ratio is found to be (0.1$-$0.5) for various LMC clouds (N\,105, N\,113, N\,159, and N\,44) by \cite{seale2012} and $\sim$ 0.2 for 30\,Doradus-10 by \cite{anderson2014} at similar spatial resolution. Thus N\,55-S clumps possess relatively high flux ratio compared to other GMCs in the LMC possibly due to the low radiation field compared to other GMCs.

There exist HCN/HCO$^{+}$ flux ratio measurements of various LMC clouds using single dish observations in the literature \citep{nishimura2016,chin1997}. These authors report the flux ratios in the range (0.5$-$0.7) for various molecular clouds of the LMC (N\,113, N\,44BC, N\,159HW, N\,214DE, N\,159W). These numbers are slightly higher than their respective values at high spatial resolution \citep[parsec scale; ][]{seale2012,anderson2014} as mentioned above. The flux ratio from single dish observations reflect the chemical composition pattern averaged over a molecular cloud scale of $10-14$ pc where the effect of local star formation activity is smeared out.  The parsec scale observations probe individual dense star forming clumps and need not be same with the flux ratio averaged over $10-14$ pc. The higher HCN/HCO+ ratios with single-dish data imply that the variation in abundance has a greater impact on the ratio than in density.  

\section{Summary}
\label{sec:summary}
We present high spatial resolution observations of HCO$^{+}$($1-0$) and HCN($1-0$) of the N\,55-S region in the LMC.  We aim to compare the the dense molecular clump properties of the N\,55-S with other active star forming regions (N\,159 and 30-Doradus) in order to understand the effect of different feedback environment on dense molecular clumps. We detect prominent HCO$^{+}(1-0)$ emission from 10 clumps and HCN($1-0$) emission from 8 clumps. Our main results are the following:

\begin{enumerate}
   \item The column density of H$_2$ gas traced by HCO$^{+}(1-0)$ emission in N\,55-S clumps are in the range $N_{\rm H_{2}} \sim$ $(0.2-1.6)$ $\times$ 10$^{22}$ cm$^{-2}$. The LTE mass and mass-surface density of the clumps are in the range $M_{\rm LTE} \sim$ (0.4$-$4.3)$\times 10^{2}$ $M_{\odot}$ and $\Sigma_{\rm p} \sim$ $(0.3-2.9)$ $\times$ 10$^{2}$ $M_{\odot}$\,pc$^{-2}$ respectively. 
    \item The volume density of H$_{2}$ gas is in the range $n_{\rm col} \sim$ ($0.3-4.9$) $\times$ 10$^{3}$ cm$^{-3}$; less than critical density of HCO$^{+}$($1-0$) emission line suggesting that the clouds are either sub-thermally excited or have very small beam filling factors. 
    \item The size-linewidth relation of HCO$^{+}$(1$-$0) clumps follow a power-law with index 0.65 $\pm$ 0.32, similar to CMZ, several extragalactic and Galactic clouds which is consistent with the $^{12}$CO clumps \citep{naslim2018}.
    \item All four identified YSOs in the N\,55-S region are in the vicinity of HCO$^{+}$(1$-$0) and HCN(1$-$0) emission peaks indicating the association of these dense clumps with recent star formation. The YSO associated clumps have relatively larger linewidths and masses compared to those without YSOs.
    \item The total H$_{2}$ mass traced by $^{12}$CO(1$-$0) and HCO$^{+}$(1$-$0) in the N\,55-S region is 2.59 $\times$ 10$^{4}$ $M_{\odot}$, and 0.70 $\times$ 10$^{3}$ $M_{\odot}$ respectively, indicating a dense gas fraction of 0.025$\pm$0.005. The dense gas fraction of N\,55-S is lower compared to other GMCs (N\,105, N\,113, N\,159, N\,44, and 30-Doradus) of the LMC indicating a relatively lower SFE. 
    \item The HCN/HCO$^{+}$ flux ratio of N\,55-S is in the range 0.46 $\pm$ 0.17 to 0.78 $\pm$ 0.12, slightly higher than the ratio (0.1 $-$ 0.5) seen in other GMCs like N\,105, N\,113, N\,159, N\,44, and 30-Doradus \citep{seale2012,anderson2014}. We interpret this to be an effect of relatively low radiation field and star formation activity in the N\,55-S. 
    \item The N\,55-S clumps possess systematically lower linewidths compared to other GMCs of the LMC \citep{seale2012,anderson2014}. We also note that the YSO associated clumps of N\,55-S show mass-surface density $\geq$ 73 $M_{\odot}$\,pc$^{-2}$ which is (6-10) times lower compared to N\,159 \citep{nayak2018} and 30-Doradus \citep{nayak2016}.  %{\bf We interpret the relatively lower linewidths and star formation mass surface density threshold of N\,55-S to be due to the quiescent environment of this GMC.}
\end{enumerate}    
    
Our study of the dense molecular clumps in N\,55-S suggests that YSOs can significantly affect the properties of dense molecular clumps by increasing the linewidths of molecular emission. Compared to other GMCs in the LMC, the dense molecular clumps of N\,55-S show smaller linewidths, lower dense gas fraction, larger HCN/HCO$^{+}$, and smaller threshold of mass surface density with YSOs indicating relatively less active star formation.

%The properties of dense molecular clumps of N\,55-S; relatively smaller line width, lower dense gas fraction, larger HCN/HCO$^{+}$, and smaller threshold of mass surface density with YSOs are all consistent with less active star formation in N\,55-S.

%\end{enumerate}

\section*{Acknowledgments}

This paper makes use of the following ALMA data:
ADS/JAO.ALMA\,$\#$2013.1.00993.S. ALMA is a partnership of the ESO, NSF,
NINS, NRC, NSC, and ASIAA. The Joint ALMA Observatory
is operated by the ESO, AUI/NRAO, and NAOJ.
This research has been supported by United Arab Emirates University, under start-up grant number 31S378. This research has been supported by the Ministry of Science and Technology of Taiwan, under grant MOST107-2119-M-001-031-MY3 and Academia Sinica under grant AS-IA-106-M03.

\clearpage


\begin{thebibliography}{}

\bibitem[Ambrocio-Cruz et al.(1998)]{ambrocio1998} Ambrocio-Cruz, P., Laval, A., Marcelin, M., et al.\ 1998, \aap, 339, 173

\bibitem[Anderson et al.(2014)]{anderson2014} Anderson, C.~N., Meier, D.~S., Ott, J., et al.\ 2014, \apj, 793, 37

\bibitem[Barnes, \& Crutcher(1990)]{barnes1990} Barnes, P.~J., \& Crutcher, R.~M.\ 1990, \apj, 351, 176

\bibitem[Barnes et al.(2011)]{barnes2011} Barnes, P.~J., Yonekura, Y., Fukui, Y., et al.\ 2011, \apjs, 196, 12

\bibitem[Bayet et al.(2011)]{bayet2011} Bayet, E., Viti, S., Hartquist, T.~W., et al.\ 2011, \mnras, 417, 627

%\bibitem[Beltr{\'a}n et al.(2006)]{beltran2006} Beltr{\'a}n, M.~T., Brand, J., Cesaroni, R., et al.\ 2006, \aap, 447, 221

\bibitem[Bolatto et al.(2008)]{bolatto2008} Bolatto, A.~D., Leroy, A.~K., Rosolowsky, E., et al.\ 2008, \apj, 686, 948

\bibitem[Bolatto et al.(2013)]{bolatto13} Bolatto, A.~D., Wolfire, M., \& Leroy, A.~K.\ 2013, \araa, 51, 207

%\bibitem[Book et al.(2009)]{book2009} Book, L.~G., Chu, Y.-H., Gruendl, R.~A., et al.\ 2009, \aj, 137, 3599

%\bibitem[Braine et al.(2017)]{braine2017} Braine, J., Shimajiri, Y., Andr{\'e}, P., et al.\ 2017, \aap, 597, A44

%\bibitem[Calzetti et al.(2007)]{calzetti2007} Calzetti, D., Kennicutt, R.~C., Engelbracht, C.~W., et al.\ 2007, \apj, 666, 870

%\bibitem[Carolan et al.(2008)]{carolan2008} Carolan, P.~B., Redman, M.~P., Keto, E., et al.\ 2008, \mnras, 383, 705

\bibitem[Carlson et al.(2012)]{carlson2012} Carlson, L.~R., Sewi{\l}o, M., Meixner, M., et al.\ 2012, \aap, 542, A66

\bibitem[Caselli et al.(2002)]{caselli2002} Caselli, P., Walmsley, C.~M., Zucconi, A., et al.\ 2002, \apj, 565, 344

\bibitem[Chen et al.(2009)]{chen2009} Chen, C.-H.~R., Chu, Y.-H., Gruendl, R.~A., et al.\ 2009, \apj, 695, 511

\bibitem[Chen et al.(2010)]{chen2010} Chen, C.-H.~R., Indebetouw, R., Chu, Y.-H., et al.\ 2010, \apj, 721, 1206

\bibitem[Chin et al.(1997)]{chin1997} Chin, Y.-N., Henkel, C., Whiteoak, J.~B., et al.\ 1997, \aap, 317, 548

%\bibitem[\protect\citeauthoryear{Dame, Hartmann \& Thaddeus}{2001}]{dame2001} Dame T.~M., Hartmann D., Thaddeus P., 2001, ApJ, 547, 792

\bibitem[Draine(1978)]{draine1978} Draine, B.~T.\ 1978, \apjs, 36, 595

\bibitem[Ellingsen et al.(2010)]{ellingsen2010} Ellingsen, S.~P., Breen, S.~L., Caswell, J.~L., et al.\ 2010, \mnras, 404, 779

\bibitem[Evans(1999)]{evans1999} Evans, N.~J.\ 1999, \araa, 37, 311

\bibitem[Fa{\'u}ndez et al.(2004)]{faundez2004} Fa{\'u}ndez, S., Bronfman, L., Garay, G., et al.\ 2004, \aap, 426, 97

%\bibitem[Federrath \& Klessen(2012)]{federrath2012} Federrath, C., \& Klessen, R.~S.\ 2012, \apj, 761, 156

\bibitem[Fontani et al.(2005)]{fontani2005} Fontani, F., Beltr{\'a}n, M.~T., Brand, J., et al.\ 2005, \aap, 432, 921

\bibitem[Fukui et al.(2015)]{fukui2015} Fukui, Y., Harada, R., Tokuda, K., et al.\ 2015, \apjl, 807, L4

\bibitem[Fukui et al.(2019)]{fukui2019} Fukui, Y., Tokuda, K., Saigo, K., et al.\ 2019, \apj, 886, 14

%\bibitem[Gao and Solomon(2004)]{gao2004}Gao, Y., and Solomon, P.M.: 2004, {\it The Astrophysical Journal} {\bf 606}, 271.

%\bibitem[Garc{\'\i}a-Burillo \emph{et al.}(2012)]{garcia12}Garc{\'\i}a-Burillo, S., Usero, A., Alonso-Herrero, A., Graci{\'a}-Carpio, J., Pereira-Santaella, M., Colina, L., Planesas, P., and Arribas, S.: 2012, {\it Astronomy and Astrophysics} {\bf 539}, A8.

\bibitem[Garrod et al.(2008)]{garrod2008} Garrod, R.~T., Widicus Weaver, S.~L., \& Herbst, E.\ 2008, \apj, 682, 283

%\bibitem[Gottlieb et al.(1975)]{gottlieb1975} Gottlieb, C.~A., Lada, C.~J., Gottlieb, E.~W., et al.\ 1975, \apj, 202, 655

\bibitem[Gruendl, \& Chu(2009)]{gruendl2009} Gruendl, R.~A., \& Chu, Y.-H.\ 2009, \apjs, 184, 172

%\bibitem[Heikkil{\"a} et al.(1999)]{heikkila1999} Heikkil{\"a}, A., Johansson, L.~E.~B., \& Olofsson, H.\ 1999, \aap, 344, 817

%\bibitem[Heaton et al.(1993)]{heaton1993} Heaton, B.~D., Little, L.~T., Yamashita, T., et al.\ 1993, \aap, 278, 238

\bibitem[Heyer et al.(2009)]{heyer2009} Heyer, M., Krawczyk, C., Duval, J., et al.\ 2009, \apj, 699, 1092

\bibitem[Hill et al.(2005)]{hill2005} Hill, T., Burton, M.~G., Minier, V., et al.\ 2005, \mnras, 363, 405

\bibitem[Hughes et al.(2010)]{hughes2010} Hughes, A., Wong, T., Ott, J., et al.\ 2010, \mnras, 406, 2065

\bibitem[Indebetouw et al.(2004)]{indebetouw2004} Indebetouw, R., Johnson, K.~E., \& Conti, P.\ 2004, \aj, 128, 2206

\bibitem[Indebetouw et al.(2013)]{indebetouw2013} Indebetouw, R., Brogan, C., Chen, C.-H.~R., et al.\ 2013, \apj, 774, 73

\bibitem[Indebetouw et al.(2020)]{indebetouw2020} Indebetouw, R., Wong, T., Chen, C.-H.~R., et al.\ 2020, \apj, 888, 56

%\bibitem[Israel et al.(2003)]{israel2003} Israel, F.~P., Johansson, L.~E.~B., Rubio, M., et al.\ 2003, \aap, 406, 817

\bibitem[\protect\citeauthoryear{Kauffmann, et al.}{2017}]{kauffmann2017} Kauffmann J., Goldsmith P.~F., Melnick G., Tolls V., Guzman A., Menten K.~M., 2017, A\&A, 605, L5

\bibitem[Kennicutt \& Evans(2012)]{kennicutt2012} Kennicutt, R.~C., \& Evans, N.~J.\ 2012, \araa, 50, 531

\bibitem[Krumholz et al.(2010)]{krumholz2010} Krumholz, M.~R., Cunningham, A.~J., Klein, R.~I., et al.\ 2010, \apj, 713, 1120

\bibitem[Krumholz et al.(2012)]{krumholz2012} Krumholz, M.~R., Dekel, A., \& McKee, C.~F.\ 2012, \apj, 745, 69

%\bibitem[Lada(1985)]{lada1985} Lada, C.~J.\ 1985, \araa, 23, 267

%\bibitem[Lada, Lombardi and Alves(2010)]{lada2010}Lada, C.J., Lombardi, M., and Alves, J.F.: 2010, {\it The Astrophysical Journal} {\bf 724}, 687.

\bibitem[Lada et al.(2012)]{lada2012} Lada, C.~J., Forbrich, J., Lombardi, M., et al.\ 2012, \apj, 745, 190

%\bibitem[Lapinov(1989)]{lapinov1989} Lapinov, A.~V.\ 1989, \sovast, 33, 132

\bibitem[Larson(1981)]{larson1981} Larson, R.~B.\ 1981, \mnras, 194, 809

\bibitem[Lee et al.(2003)]{lee2003} Lee, J.-E., Evans, N.~J., Shirley, Y.~L., et al.\ 2003, \apj, 583, 789

\bibitem[Leroy et al.(2011)]{leroy2011} Leroy, A.~K., Bolatto, A., Gordon, K., et al.\ 2011, \apj, 737, 12

%\bibitem[Leroy et al.(2015)]{leroy2015} Leroy, A.~K., Bolatto, A.~D., Ostriker, E.~C., et al.\ 2015, \apj, 801, 25

%\bibitem[Leroy et al.(2017)]{leroy2017} Leroy, A.~K., Usero, A., Schruba, A., et al.\ 2017, \apj, 835, 217

\bibitem[Loren et al.(1990)]{loren1990} Loren, R.~B., Wootten, A., \& Wilking, B.~A.\ 1990, \apj, 365, 269

%\bibitem[Loughnane et al.(2012)]{loughnane2012} Loughnane, R.~M., Redman, M.~P., Thompson, M.~A., et al.\ 2012, \mnras, 420, 1367

\bibitem[Mangum \& Shirley(2015)]{mangum2015} Mangum, J.~G. \& Shirley, Y.~L.\ 2015, \pasp, 127, 266


%\bibitem[Mardones et al.(1997)]{mardones1997} Mardones, D., Myers, P.~C., Tafalla, M., et al.\ 1997, \apj, 489, 719

\bibitem[McMullin et al.(2007)]{mcmullin2007} McMullin, J.~P., Waters, B., Schiebel, D., et al.\ 2007, Astronomical Data Analysis Software and Systems XVI, 127

%\bibitem[Meixner et al.(2006)]{meixner2006} Meixner, M., Gordon, K.~D., Indebetouw, R., et al.\ 2006, \aj, 132, 2268
\bibitem[Meijerink et al.(2011)]{meijerink2011} Meijerink, R., Spaans, M., Loenen, A.~F., et al.\ 2011, \aap, 525, A119

\bibitem[Meixner et al.(2010)]{meixner2010} Meixner, M., Galliano, F., Hony, S., et al.\ 2010, \aap, 518, L71


%\bibitem[Millar \& Herbst(1990)]{millar1990} Millar, T.~J., \& Herbst, E.\ 1990, \mnras, 242, 92

%\bibitem[Miyamoto et al.(2017)]{miyamoto2017} Miyamoto, Y., Nakai, N., Seta, M., et al.\ 2017, \pasj, 69, 83

%\bibitem[Mizuno et al.(2010)]{mizuno2010} Mizuno, Y., Kawamura, A., Onishi, T., et al.\ 2010, \pasj, 62, 51

\bibitem[Muller et al.(2010)]{muller2010} Muller, E., Ott, J., Hughes, A., et al.\ 2010, \apj, 712, 1248

%\bibitem[Myers(1983)]{myers1983} Myers, P.~C.\ 1983, \apj, 270, 105

%\bibitem[Myers(1987)]{myers1987} Myers, P.~C.\ 1987, Star Forming Regions, 33

\bibitem[Naslim et al.(2015)]{naslim2015} Naslim, N., Kemper, F., Madden, S.~C., et al.\ 2015, \mnras, 446, 2490

\bibitem[Naslim et al.(2018)]{naslim2018} Naslim, N., Tokuda, K., Onishi, T., et al.\ 2018, \apj, 853, 175

\bibitem[Nayak et al.(2016)]{nayak2016} Nayak, O., Meixner, M., Indebetouw, R., et al.\ 2016, \apj, 831, 32

\bibitem[Nayak et al.(2018)]{nayak2018} Nayak, O., Meixner, M., Fukui, Y., et al.\ 2018, \apj, 854, 154


\bibitem[Nishimura et al.(2016)]{nishimura2016} Nishimura, Y., Shimonishi, T., Watanabe, Y., et al.\ 2016, \apj, 818, 161


%\bibitem[Ntormousi et al.(2011)]{ntormousi2011} Ntormousi, E., Burkert, A., Fierlinger, K., et al.\ 2011, \apj, 731, 13

\bibitem[Ochsendorf et al.(2017)]{ochsendorf2017} Ochsendorf, B.~B., Meixner, M., Roman-Duval, J., et al.\ 2017, \apj, 841, 109

\bibitem[Olsen et al.(2001)]{olsen2001} Olsen, K.~A.~G., Kim, S., \& Buss, J.~F.\ 2001, \aj, 121, 3075

%\bibitem[Papadopoulos(2007)]{papadopoulos2007} Papadopoulos, P.~P.\ 2007, \apj, 656, 792

\bibitem[Pietrzy{\'n}ski et al.(2019)]{pietrzynski2019} Pietrzy{\'n}ski, G., Graczyk, D., Gallenne, A., et al.\ 2019, \nat, 567, 200

\bibitem[Pineda et al.(2009)]{pineda2009} Pineda, J.~L., Ott, J., Klein, U., et al.\ 2009, \apj, 703, 736

\bibitem[Pineda et al.(2017)]{pineda2017} Pineda, J.~L., Langer, W.~D., Goldsmith, P.~F., et al.\ 2017, \apj, 839, 107


\bibitem[Querejeta et al.(2019)]{querejeta2019} Querejeta, M., Schinnerer, E., Schruba, A., et al.\ 2019, \aap, 625, A19


%\bibitem[Redman et al.(2006)]{redman2006} Redman, M.~P., Keto, E., \& Rawlings, J.~M.~C.\ 2006, \mnras, 370, L1

\bibitem[Retes-Romero et al.(2017)]{romero2017} Retes-Romero, R., Mayya, Y.~D., Luna, A., et al.\ 2017, \apj, 839, 113

\bibitem[Rohlfs, \& Wilson(2004)]{rohlfs2004} Rohlfs, K., \& Wilson, T.~L.\ 2004, Tools of radio astronomy

\bibitem[Rosolowsky et al.(2008)]{rosolowsky2008} Rosolowsky, E.~W., Pineda, J.~E., Kauffmann, J., et al.\ 2008, \apj, 679, 1338

%\bibitem[Saigo et al.(2017)]{saigo2017} Saigo, K., Onishi, T., Nayak, O., et al.\ 2017, \apj, 835, 108

\bibitem[Saito et al.(2006)]{saito2006} Saito, H., Saito, M., Moriguchi, Y., et al.\ 2006, \pasj, 58, 343

%\bibitem[Sch{\"o}ier et al.(2005)]{schoier2005} Sch{\"o}ier, F.~L., van der Tak, F.~F.~S., van Dishoeck, E.~F., et al.\ 2005, \aap, 432, 369

%\bibitem[Schirmacher \& Winter(1993)]{schirmacher1993} Schirmacher, A., \& Winter, H.\ 1993, \pra, 47, 4891

\bibitem[Seale et al.(2012)]{seale2012} Seale, J.~P., Looney, L.~W., Wong, T., et al.\ 2012, \apj, 751, 42

\bibitem[Seale et al.(2014)]{seale2014} Seale, J.~P., Meixner, M., Sewi{\l}o, M., et al.\ 2014, \aj, 148, 124

\bibitem[Shetty et al.(2012)]{shetty2012} Shetty, R., Beaumont, C.~N., Burton, M.~G., et al.\ 2012, \mnras, 425, 720

%\bibitem[Shimajiri et al.(2017)]{shimajiri2017}Shimajiri, Y., Andr{\'e}, P., Braine, J., K{\"o}nyves, V., Schneider, N., Bontemps, S., Ladjelate, B., Roy, A., Gao, Y., and Chen, H.: 2017, {\it Astronomy and Astrophysics} {\bf 604}, A74.

\bibitem[\protect\citeauthoryear{Shirley}{2015}]{shirley2015} Shirley Y.~L., 2015, PASP, 127, 299

\bibitem[Smith \& MCELS Team(1998)]{smith1998} Smith, R.~C., \& MCELS Team\ 1998, \pasa, 15, 163

\bibitem[Solomon et al.(1987)]{solomon1987} Solomon, P.~M., Rivolo, A.~R., Barrett, J., et al.\ 1987, \apj, 319, 730

\bibitem[Strong et al.(1988)]{strong1988} Strong, A.~W., Bloemen, J.~B.~G.~M., Dame, T.~M., et al.\ 1988, \aap, 207, 1

%\bibitem[Stutzki, \& Guesten(1990)]{stutzki1990} Stutzki, J., \& Guesten, R.\ 1990, \apj, 356, 513

\bibitem[Tokuda et al.(2019)]{tokuda2019} Tokuda, K., Fukui, Y., Harada, R., et al.\ 2019, \apj, 886, 15

\bibitem[van der Tak et al.(2000)]{vandertak2000} van der Tak, F.~F.~S., van Dishoeck, E.~F., Evans, N.~J., et al.\ 2000, \apj, 537, 283

\bibitem[Whitney et al.(2008)]{whitney2008} Whitney, B.~A., Sewilo, M., Indebetouw, R., et al.\ 2008, \aj, 136, 18

\bibitem[Williams et al.(2000)]{williams2000} Williams, J.~P., Blitz, L., \& McKee, C.~F.\ 2000, Protostars and Planets IV, 97

\bibitem[Wong et al.(2006)]{wong2006} Wong, T., Whiteoak, J.~B., Ott, J., et al.\ 2006, \apj, 649, 224

\bibitem[Wong et al.(2011)]{wong2011} Wong, T., Hughes, A., Ott, J., et al.\ 2011, \apjs, 197, 16

%\bibitem[Wu \emph{et al.}(2005)]{wu05}Wu, J., Evans, N.J., Gao, Y., Solomon, P.M., Shirley, Y.L., and Vanden Bout, P.A.: 2005, {\it The Astrophysical Journal} {\bf 635}, L173.

\bibitem[Yamaguchi et al.(2001)]{yamaguchi2001} Yamaguchi, R., Mizuno, N., Onishi, T., et al.\ 2001, \pasj, 53, 959

%\bibitem[Zhou et al.(1989)]{zhou1989} Zhou, S., Wu, Y., Evans, N.~J., et al.\ 1989, \apj, 346, 168

\bibitem[Zinchenko et al.(2009)]{zinchenko2009} Zinchenko, I., Caselli, P., \& Pirogov, L.\ 2009, \mnras, 395, 2234

\end{thebibliography}
\end{document}